\documentclass[pdflatex,sn-mathphys-num]{sn-jnl}

\usepackage{tikz}
\usetikzlibrary{shapes.geometric, arrows}

\tikzset{
    block/.style = {draw, rectangle, align=center, minimum height=2em, minimum width=4em},
    decision/.style = {diamond, draw, aspect=2, align=center, inner sep=1pt}
}

\usepackage{tabularx}
\usepackage{graphicx}%
\usepackage{multirow}%
\usepackage{amsmath,amssymb,amsfonts}%
\usepackage{amsthm}%
\usepackage{mathrsfs}%
\usepackage[title]{appendix}%
\usepackage{xcolor}%
\usepackage{textcomp}%
\usepackage{manyfoot}%
\usepackage{booktabs}%
\usepackage{algorithm}%
\usepackage{algorithmicx}%
\usepackage{algpseudocode}%
\usepackage{listings}%
\usepackage{adjustbox}%

\usepackage{caption}
\usepackage{lineno,hyperref}
\usepackage{tablefootnote}
\usepackage{subcaption}
\usepackage{soul,color}
\usepackage{float}
\usepackage{comment}
\usepackage{dblfloatfix}
\usepackage{bbding}
\usepackage{booktabs}  

\theoremstyle{thmstyleone}%
\theoremstyle{thmstyletwo}%

\theoremstyle{thmstylethree}%

\newcolumntype{L}{>{\raggedright\arraybackslash}p{3.5cm}} 
\newcolumntype{C}{>{\centering\arraybackslash}X} 

\begin{document}

\title[Adv-IOT]{Targeted Adversarial Traffic Generation : Black-box Approach to Evade Intrusion Detection Systems in IoT Networks}


\author*{\fnm{Islam} \sur{Debicha}}\email{debichasislam@gmail.com}

\author{\fnm{Tayeb} \sur{Kenaza}}

\author{\fnm{Ishak} \sur{Charfi}}

\author{\fnm{Salah} \sur{Mosbah}}

\author{\fnm{Mehdi} \sur{Sehaki}}

\author{\fnm{Jean-Michel} \sur{Dricot}}

\affil{\orgdiv{Computer Security Laboratory}, \orgname{Ecole Militaire Polytechnique}, \orgaddress{
\country{Algeria}}}

\affil{\orgdiv{Cybersecurity Research Center}, \orgname{Université Libre de Bruxelles}, \orgaddress{
\country{Belgium}}}


\abstract{The integration of machine learning (ML) algorithms into Internet of Things (IoT) applications has introduced significant advantages alongside vulnerabilities to adversarial attacks, especially within IoT-based intrusion detection systems (IDS). While theoretical adversarial attacks have been extensively studied, practical implementation constraints have often been overlooked. This research addresses this gap by evaluating the feasibility of evasion attacks on IoT network-based IDSs,  employing a novel black-box adversarial attack. Our study aims to bridge theoretical vulnerabilities with real-world applicability, enhancing understanding and defense against sophisticated threats in modern IoT ecosystems. Additionally, we propose a defense scheme tailored to mitigate the impact of evasion attacks, thereby reinforcing the resilience of ML-based IDSs. Our findings demonstrate successful evasion attacks against IDSs, underscoring their susceptibility to advanced techniques. In contrast, we proposed a defense mechanism that exhibits robust performance by effectively detecting the majority of adversarial traffic, showcasing promising outcomes compared to current state-of-the-art defenses. By addressing these critical cybersecurity challenges, our research contributes to advancing IoT security and provides insights for developing more resilient IDS.}




\maketitle

\section{Introduction}
\par{The proliferation of Internet of Things (IoT) technologies has transformed numerous sectors, including smart cities, healthcare, and industrial automation~\cite{heidari2023internet}. \textcolor{black}{However, despite their potential, IoT networks are inherently vulnerable due to decentralized architectures and limited resources~\cite{rahman2020internet}.}}

\par{Network Intrusion Detection Systems (NIDSs) play a pivotal role in safeguarding IoT networks by monitoring traffic and identifying malicious activities. To address the limitations of traditional signature-based NIDSs, machine learning (ML) algorithms have been increasingly incorporated, thereby enhancing their ability to detect both known and emerging threats~\cite{debicha2023adv}. \textcolor{black}{Nevertheless, recent studies have revealed that these ML-based IDSs are susceptible to adversarial attacks, where minimal, carefully crafted perturbations can lead to misclassification errors~\cite{he2020towards,ilyas2019adversarial, debicha2023tad}.}}

\par{While adversarial attacks have been extensively investigated in domains such as computer vision~\cite{cai2022zero,wei2022towards} and natural language processing (NLP)~\cite{zang2019word}, their application to cybersecurity remains underexplored~\cite{apruzzese2019evaluating}. \textcolor{black}{This research gap is particularly critical for ML-based IDSs in IoT networks, where the unique characteristics of network traffic demand tailored adversarial strategies.}}

\par{Unlike deep learning (DL) approaches, ML models offer advantages in terms of computational efficiency and performance on tabular data, making them well-suited for resource-constrained IoT environments~\cite{mccarthy2022functionality,martins2020adversarial}. However, the vulnerability of these models to adversarial attacks highlights the need for robust defenses that consider domain-specific constraints.}

\par{The design of practical adversarial attacks against NIDSs necessitates the consideration of three critical factors: knowledge restriction, domain constraints, and manipulation space~\cite{debicha2023review}. \textcolor{black}{In our work, we address these factors by proposing a black-box adversarial attack - Distance to Target Center (D2TC) - that effectively manipulates raw network traffic while preserving its semantic integrity.}}

\par{To the best of our knowledge, this study is the first to propose a realistic black-box adversarial attack that integrates domain constraints and manipulation space considerations against NIDSs in IoT networks. The primary contributions of this work are as follows:}
\begin{itemize}
\item First, it presents a comprehensive examination of the constraints necessary for generating effective adversarial perturbations while preserving the underlying functionality of IoT attack traffic. This analysis emphasizes the importance of adhering to semantic and syntactic constraints during the generation process.
\item Second, it introduces a novel black-box adversarial attack, termed Distance to Target Center (D2TC), specifically designed to generate valid adversarial traffic targeting a particular class of IoT traffic. The D2TC attack is demonstrated to effectively evade machine learning-based intrusion detection systems.
\item Third, it proposes a robust defense mechanism against the D2TC attack, leveraging an adversarial detection strategy. This mechanism employs machine learning algorithms to detect and classify adversarial traffic, enabling the intrusion detection system to accurately identify and mitigate D2TC-generated attacks.
\end{itemize}

\par{\textcolor{black}{The remainder of this paper is organized as follows. Section~\ref{sec:back} reviews related work in adversarial attacks and defenses for IDSs in IoT networks. Section~\ref{sec:proposed} details our proposed D2TC attack method. Section~\ref{sec:def} introduces our defense mechanism. Section~\ref{sec:results} presents the experimental setup, results, and analysis. Finally, Section~\ref{sec:conclusion} concludes the paper and outlines directions for future research.}}

\section{Background \& Related works }
\label{sec:back}
\subsection{Internet Of Things}
\par{The Internet of Things (IoT) is a framework in which all objects have a representation and presence on the Internet. More specifically, the IoT aims to provide new applications and services that bridge the physical and virtual worlds. Machine-to-Machine (M2M) communications form the foundational layer, enabling interactions between devices ("Things") and applications hosted in the cloud~\cite{mouha2021internet}.There are many applications of IoT, including Industry 4.0, smart homes, smart cities, healthcare systems, the automotive sector, public services, and critical infrastructure~\cite{9698229,DAS20191,Jeyaraj2019,
tajdini2020wireless,9666894}. These applications use the connection between IoT devices to increase efficiency, support better decisions, and deliver real-time insights across various domains~\cite{laghari2021review}.However, IoT devices are often vulnerable to threats such as weak authentication mechanisms, insecure data storage, and communication protocols, which make them targets for cyberattacks, including unauthorized access, data breaches, and distributed denial-of-service (DDoS) attacks~\cite{baho2023analysis}. These vulnerabilities highlight the necessity of deploying robust intrusion detection mechanisms tailored to IoT environments~\cite{khraisat2021critical}.}
\subsection{Intrusion Detection System}
\par{An Intrusion Detection System (IDS) is software or a device that detects attacks on a system and notifies the system administrator. It can monitor an individual system or perform local analysis in a network to identify potential threats. IDS provides three key security services: (i) ensuring data confidentiality by securing stored data, (ii) ensuring data availability by allowing access only to authorized users, and (iii) ensuring data integrity by maintaining the accuracy and consistency of the data within the system~\cite{al2025deep}. Based on detection location, IDS can be classified as: Host-based IDS (HIDS), which monitors activities on individual devices, and Network-based IDS (NIDS), which analyzes traffic across a network to detect threats. Based on the detection method, IDS can be categorized as: Signature-based IDS (SIDS), which identifies threats using known attack patterns, and Anomaly-based IDS (AIDS), which detects deviations from normal behavior to identify potential threats~\cite{khraisat2021critical}. While SIDS provides excellent detection accuracy for previously known intrusions, it is ineffective against zero-day attacks due to the absence of corresponding signatures in its database. As a result, SIDS is less commonly used, while AIDS, which often uses machine learning (ML) techniques, has become more prevalent for its ability to identify unknown threats by detecting deviations from normal behavior~\cite{khraisat2021critical}.}

\subsection{Adversarial Attacks}

Adversarial attacks are deliberate attempts to exploit vulnerabilities in ML models by introducing carefully crafted inputs designed to deceive the model~\cite{zhou2022adversarial}. These attacks can be categorized along several dimensions:

\begin{itemize}
\item \textbf{Adversarial goal}: Adversarial attacks are typically classified into untargeted and targeted types. \textit{Untargeted attacks} aim to cause the model to make any incorrect prediction, prioritizing misclassification without focusing on a specific outcome. On the other hand, \textit{targeted attacks} are more precise, seeking to manipulate the model's prediction toward a specific class chosen by the attacker~\cite{debicha2023review}.
\item \textbf{Adversarial capability}: The effectiveness of an adversarial attack often depends on the attacker's level of access or knowledge about the target system. In \textit{white-box attacks}, the attacker has full access to the model's architecture, parameters, and training data. \textit{Black-box attacks}, in contrast, assume no direct access to the model, relying solely on input-output interactions to craft adversarial examples. \textit{Grey-box attacks} fall between these extremes, where the attacker has partial knowledge, such as access to the architecture but not the training data~\cite{debicha2023adv}.
\item \textbf{Attack phase}: Adversarial attacks can occur at different stages of the machine learning pipeline. \textit{Poisoning attacks} target the training phase by injecting malicious samples into the dataset, causing the model to learn incorrect patterns. Conversely, \textit{evasion attacks} are executed during the inference phase, where attackers add imperceptible perturbations to test samples to fool the trained model without altering the training process~\cite{zhou2022adversarial}.
\end{itemize}

\subsection{Related Works}
\label{sec:RelatedWorks}

Several recent studies have investigated the use of machine learning (ML) techniques in designing intrusion detection systems (IDSs) for IoT networks. These studies focus on improving detection accuracy, optimizing efficiency for resource-constrained devices, and enhancing adaptability to evolving threats. Notable advancements include feature selection methods, anomaly-based detection mechanisms, and novel frameworks addressing IoT-specific challenges~\cite{islam2021towards,fernando2023enhancing,sarhan2024feature,altulaihan2024anomaly}.

While significant progress has been made in integrating ML techniques into IDSs for IoT networks, ensuring their robustness remains a critical challenge. Numerous studies have demonstrated that ML-based IDSs are susceptible to adversarial attacks, where attackers introduce subtle perturbations to input data to manipulate model predictions~\cite{zhao2021attackgan,dong2025masqueradegan,alslman2024robust,yuan2024simple,barik2024adversarial}. Among these, evasion attacks are particularly concerning due to their practicality during deployment~\cite{debicha2023adv}. For instance, poisoning attacks require access to training data, which is often restricted in real-world scenarios, making evasion attacks more feasible~\cite{apruzzese2022modeling}.

\textcolor{black}{
In addition to our focus on adversarial attacks targeting IDS in IoT networks, recent advances in adversarial robustness across other domains offer valuable insights. For instance, in computer vision and natural language processing, techniques such as the Fast Gradient Sign Method (FGSM) have been applied to generate adversarial point clouds, revealing vulnerabilities in models like PointNet~\cite{zhang2019defense}. Furthermore, recent work in image processing has demonstrated that leveraging edge feature information can enhance robustness against adversarial attacks~\cite{xiao2023defed}, while the innovative "attack as defense" framework evaluates an input's robustness based on the cost of attacking it, providing a novel detection mechanism~\cite{zhao2024attack}. In parallel, cutting-edge research in DL-based traffic systems - such as Hierarchical Graph Convolution Networks for Traffic Forecasting~\cite{guo2021hierarchical} and Multi-range Spatial-temporal Transformer Networks for Traffic Forecast via Structural Entropy Optimization~\cite{zou2024multispans} - has shown effective methods for capturing complex traffic dynamics. Similarly, advanced anomaly detection approaches, including the Se-GSL framework~\cite{zou2023se} and the Random Partitioning Forest for network intrusion detection~\cite{marteau2021random}, illustrate innovative techniques for identifying subtle irregularities in network behavior. Although these techniques are developed in non-IoT contexts, they offer promising methodologies that could be adapted to further enhance the adversarial robustness of IoT-based IDSs. Future research should explore the integration of these cross-domain insights to develop more comprehensive and resilient defense strategies.
}

Furthermore, Deep learning (DL)-based IDSs are often preferred due to their ability to learn complex representations. However, they come with significant drawbacks, including higher vulnerability to adversarial attacks, greater computational demands, and dependency on large amounts of labeled data. In contrast, ML models such as random forests and XGBoost offer better accuracy on tabular data, which is crucial in intrusion detection~\cite{shwartz2022tabular}. Additionally, their lower computational footprint makes them well-suited for IoT environments, where energy efficiency and low latency are critical~\cite{islam2021towards}. Yet, adversarial attacks targeting ML models remain significantly less studied compared to DL architectures, justifying our focus on these models.

McCarthy et al. \cite{mccarthy2022functionality} highlight that while adversarial attacks on DL models are well-documented, ML models remain underexplored despite their widespread adoption in cybersecurity applications. Similarly, \cite{martins2020adversarial} provide a systematic review showing that most adversarial attack research has been conducted in computer vision and natural language processing, with significantly fewer studies targeting ML-based IDS in network security. \textcolor{black}{Table~\ref{tab:related_work} summarizes key contributions in this area.}
\begin{table}[htbp]
\centering
\caption{Summary of Recent Works on ML-based IDS}
\label{tab:related_work}
\begin{tabular}{lll}
\toprule
\textbf{Study} & \textbf{Contribution} & \textbf{Limitations} \\
\midrule
Islam et al.~\cite{islam2021towards} & Feature selection for lightweight IDS & Limited to specific IoT protocols \\
Sarhan et al.~\cite{sarhan2024feature} & Novel framework for IoT-specific challenges & Requires large labeled datasets \\
Altulaihan et al.~\cite{altulaihan2024anomaly} & Real-time anomaly detection & Susceptible to adversarial attacks \\
Alslman et al.~\cite{alslman2024robust} & Robust IDS using adversarial training & High resource requirements \\
Yuan et al.~\cite{yuan2024simple} & Simple yet effective adversarial defenses & Limited evaluation on IoT datasets \\
Barik et al.~\cite{barik2024adversarial} & Adversarial robustness for tabular data & Not IoT-specific \\
\bottomrule
\end{tabular}
\end{table}

By focusing exclusively on ML-based IDS, this research aims to bridge this gap by evaluating their robustness against adversarial threats and developing specific defense mechanisms tailored to their characteristics.

\section{Black-box Evasion Attack}
\label{sec:proposed}

\subsection{Threat Model}
\label{subsec:threatsceanario}

\begin{figure*}[h]
\centering
\includegraphics[width=\columnwidth]{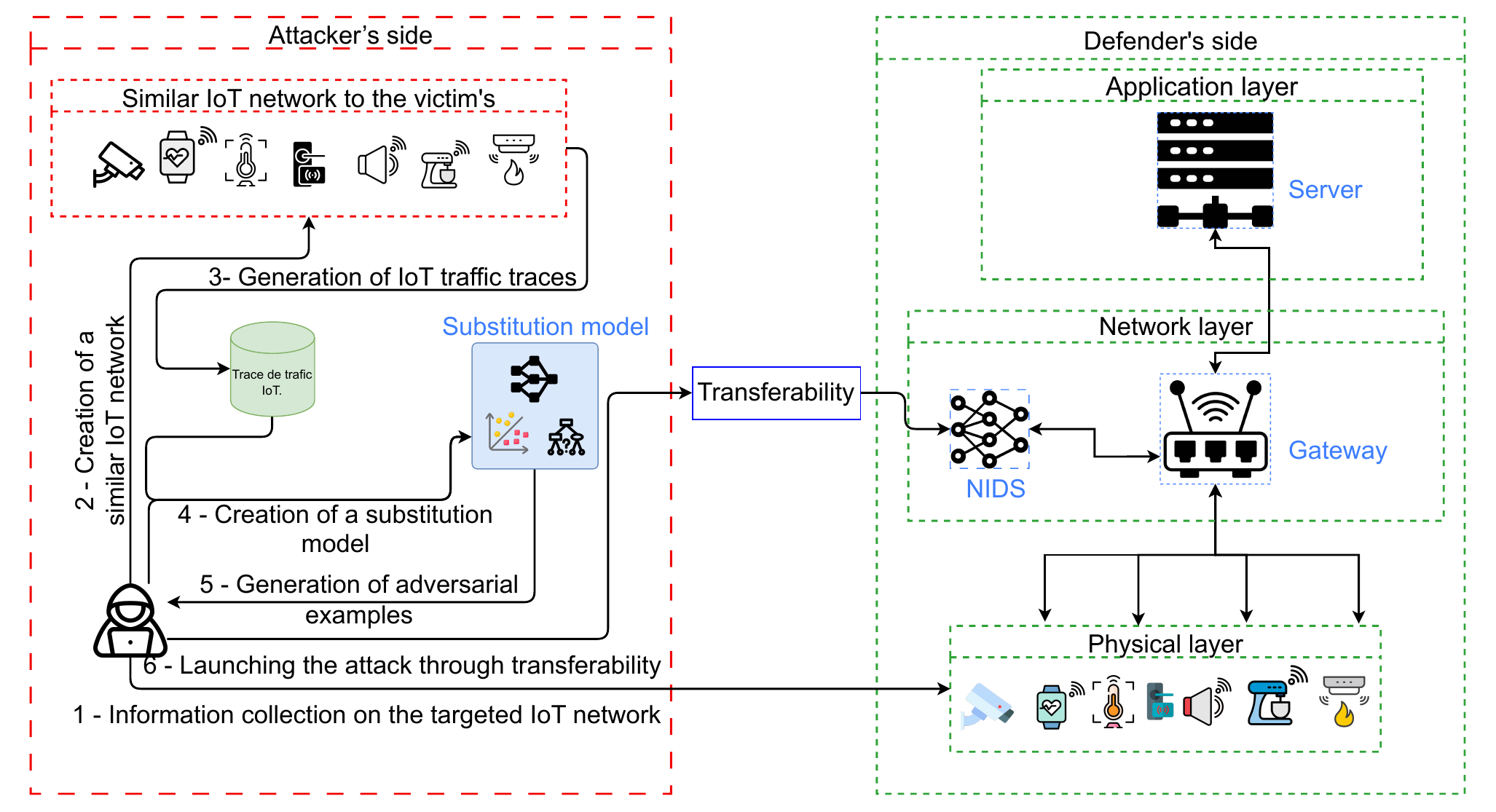}
\caption{Schematic illustration of the threat scenario, showing how the attacker replicates a substitute  IoT environment, builds a substitution model, and leverages transferability to craft adversarial examples that evade the defender’s IoT-based NIDS.}
\label{fig:threatsenario}
\end{figure*}

Figure~\ref{fig:threatsenario} schematically illustrates our threat scenario. In this setup, the defender’s network employs an IoT-based NIDS that uses machine learning algorithms to detect attacks. Prior to real-time monitoring, the NIDS is trained offline on historical data and network patterns, and operates as a flow-based system to efficiently process high-speed network traffic. A flow exporter extracts key network features from each flow, which are then preprocessed and fed into the NIDS for classification.

\color{black}
In our threat scenario, the attacker is assumed to be an outsider operating under a black-box setting. This means that the attacker has limited knowledge of the target IoT network and no direct access to the IDS model used by the defender, including its architecture, parameters, and training data. Instead, the attacker gathers publicly available information and constructs a substitute  dataset that closely mimics the network traffic characteristics of the target IoT environment, while remaining distinct from the defender’s data. Using this substitute  dataset, the attacker builds a substitution model that approximates the behavior of the defender’s IDS. The attacker then leverages this substitution model to generate adversarial network traffic via the D2TC method. Thanks to the transferability property, this adversarial traffic, crafted on the substitute  model, remains effective against the defender's IDS when injected at inference time. Our network setup is designed to reflect realistic IoT conditions, ensuring that the substitution model is a close approximation of the actual operational environment and that the overall threat scenario is both comprehensive and representative of real-world attacks.
\color{black}
As illustrated in Figure~\ref{fig:threatsenario}, the overall attack scenario unfolds in six sequential steps:
\begin{enumerate}
    \item \textbf{Information Collection:} The attacker gathers data on the target IoT network’s structure and behavior from public sources.
    \item \textbf{Network Replication:} Based on the collected data, the attacker replicates a similar IoT environment (e.g., using commercially available systems typical of Industrial or Smart Home IoT).
    \item \textbf{Traffic Trace Generation:} The attacker generates IoT traffic traces that mimic normal network activity.
    \item \textbf{Substitution Model Creation:} Using the substitute  dataset, a substitution model is developed to approximate the defender’s IDS behavior.
    \item \textbf{Adversarial Example Generation:} The D2TC method is employed on the substitution model to craft adversarial examples that aim to bypass the IDS.
    \item \textbf{Attack Launch via Transferability:} The adversarial examples, benefiting from the transferability property, are injected into the target network to evade detection.
\end{enumerate}

In accordance with our prior research on NIDS evasion~\cite{debicha2023adv}, the attacker's objectives, knowledge, and capabilities are defined as follows:
\begin{itemize}
    \item \textbf{Objective:} The attacker pursues two main objectives: to exploit security vulnerabilities in the network to carry out a malicious attack, and to manipulate the target NIDS so that it misclassifies the attack as benign.
    \color{black}
     \item \textbf{Knowledge:} As an outsider, the attacker lacks direct access to the enterprise's internal network and the defender’s IDS configurations. Instead, the attacker relies on remote network scanning, passive monitoring of publicly accessible network interfaces, and open-source intelligence to infer general network characteristics and identify IoT devices. This external approach provides only limited, approximate information about the target environment.
     \color{black}
    \item \textbf{Capabilities:} The attacker is capable of replicating a simulated IoT network that approximates the victim’s environment, enabling the generation of a substitute  dataset for building a substitution model that mimics the defender’s IDS behavior.
\end{itemize}

\textcolor{black}{
Concerning the flow exporter, an attacker has the ability to manipulate elements within the traffic domain without possessing detailed knowledge of the specific features utilized by the NIDS. This capability arises from the widespread use of common network metrics - such as \textit{duration}, \textit{packet count}, and \textit{size} - in NIDS. Attackers may assume that these metrics form part of the defender's feature set. These features can be obtained from well-known exporters like Argus, CIC-FlowMeter, or nProbe, as well as from scholarly articles~\cite{sarhan2022towards, pektacs2017effective} that discuss the features used by NIDS models.}

This detailed threat model reflects a realistic black-box attack scenario where the attacker, acting as an outsider, leverages publicly available information and substitute  data to craft effective adversarial examples against the defender's IDS.

\subsection{Network Traffic Manipulation}
\label{sec:f.manipulation}

To effectively generate adversarial examples, attackers must manipulate key network features while preserving the semantic integrity of the traffic. Based on the attacker's ability to modify network features, these features are classified into three distinct groups:

\begin{enumerate}
    \item \textbf{First group:} Features that can be directly manipulated by an attacker. This group includes the features listed in Table~\ref{tab:feature_modifiable}.
        
    \begin{table}[h!]
        \centering
        \caption{Description of Common Modifiable Features}
        \label{tab:feature_modifiable}
        \begin{tabularx}{0.8\textwidth}{L X}
            \toprule
            \textbf{Features} & \textbf{Description} \\
            \midrule
            Dur    & Duration of the flow \\
            spkts  & Number of packets from source to destination \\ 
            sbytes & Number of bytes from source to destination \\ 
            \bottomrule
        \end{tabularx}
    \end{table}

    \item \textbf{Second group:} Features that are dependent on the first group and can be indirectly manipulated. Table~\ref{tab:feature_dependent} illustrates these features and their dependencies.
    
    \begin{table}[h!]
        \centering
        \caption{Description of Common Dependent Features}
        \label{tab:feature_dependent}
        \begin{tabularx}{0.8\textwidth}{L L L}
            \toprule
            \textbf{Features} & \textbf{Description} & \textbf{Dependency} \\
            \midrule
            Pkts  & Total number of packets & Spkts, Dpkts \\
            Bytes & Total number of bytes & Sbytes, Dbytes \\ 
            Rate  & Total number of packets per second & Dur, Pkts \\ 
            Srate & Packets from source to destination per second  & Dur, Spkts \\ 
            Drate & Packets from destination to source per second & Dur, Dpkts \\ 
            \bottomrule
        \end{tabularx}
    \end{table}
    
    \item \textbf{Third group:} Features that are unmodifiable and cannot be altered by the attacker, as summarized in Table~\ref{tab:feature_non_modifiable}.
    
    \begin{table}[h!]
        \centering
        \caption{Description of Common Non-Modifiable Features}
        \label{tab:feature_non_modifiable}
        \begin{tabularx}{0.8\textwidth}{L X}
            \toprule
            \textbf{Features} & \textbf{Description} \\
            \midrule
            Dpkts  & Number of packets from destination to source \\ 
            Dbytes & Number of bytes from destination to source \\ 
            Proto  & Textual representation of protocols  \\
            Sport  & Source port number \\ 
            Dport  & Destination port number  \\ 
            \bottomrule
        \end{tabularx}
    \end{table}
\end{enumerate}

To ensure that malicious communications remain effective while avoiding detection, attackers must manipulate these network features in a manner that adheres to the syntactic and semantic constraints of the underlying protocols. This adherence is critical to preserving the functional integrity of the traffic while evading IDS detection.

Attackers can employ several strategies to manipulate these features:

\color{black}
\begin{enumerate}
    \item \textbf{Time Interval Adjustment:}  
    As an outsider, the attacker cannot modify internal system components such as the flow exporter or IDS. Instead, the attacker manipulates the inter-packet duration by controlling the timing of packet transmissions at the application level using packet injection tools (e.g., Scapy) or custom scripts. For example, if an IoT device typically sends packets every 1 second, the attacker might introduce a slight delay, increasing the interval to 1.1 seconds. This change directly affects the \emph{Duration} feature, leading to a longer flow duration and indirectly impacting derived metrics such as \emph{Rate}, \emph{Srate}, and \emph{Drate}. Importantly, as the packet contents remain unaltered, this strategy maintains protocol compliance while subtly modifying timing characteristics.
    
    \item \textbf{Packet Injection/Withholding:}  
    This strategy targets the packet count. The attacker may inject an additional packet into a flow, thereby increasing the \emph{Spkts} (source packet count), or withhold a packet, reducing the count. Such modifications directly influence the \emph{Spkts} feature and indirectly affect the overall packet count (\emph{Pkts}) as well as derived metrics such as \emph{Rate} and \emph{Srate}. For instance, in a flow originally containing 10 packets, injecting one extra packet disrupts the expected count and can mislead the IDS.
    
    \item \textbf{Byte Count Manipulation:}  
    Here, the attacker targets the \emph{Sbytes} feature by altering the payload size. For example, if an IoT device typically transmits 500 bytes per packet, the attacker might add 20 bytes of padding (byte expansion) to produce 520 bytes per packet. Alternatively, byte compression may reduce the transmitted size. These adjustments directly affect the \emph{Sbytes} metric and indirectly modify the overall byte count (\emph{Bytes}), thereby controlling the perceived data volume.
\end{enumerate}

By employing these strategies, the attacker is able to subtly modify key network features while preserving the overall structure and semantic integrity of the traffic. This balance ensures that adversarial examples remain valid under protocol constraints, making the attack both effective and difficult to detect.

\color{black}

\subsection{Adversarial Instances Generation}
\label{sec:AdvInsGen}
The Distance to Target Center (D2TC) method provides an intuitive approach for generating adversarial perturbations. It begins by computing the mean value of each network feature from benign instances and then adds to a malicious instance a perturbation equal to a portion of the difference between this benign mean and the malicious value. This perturbation is subsequently projected into the space of valid values, ensuring that the generated adversarial instance remains plausible.

\begin{figure*}[h!]
\centering
\includegraphics[scale=0.8]{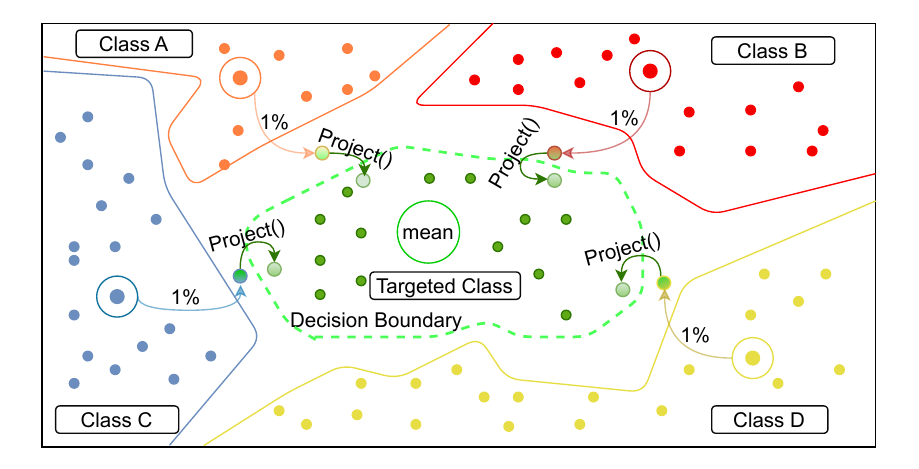}
\caption{Conceptual illustration of the D2TC approach, where malicious instances are iteratively shifted toward the benign mean and projected into valid ranges. Once inside the benign region, the instance is misclassified as benign, forming an adversarial example}
\label{fig:advtransfo}
\end{figure*}

Figure~\ref{fig:advtransfo} illustrates the transformation process. As perturbations are applied, the malicious instance gradually shifts toward the benign mean. Once the instance falls within acceptable bounds (as enforced by the projection), it is misclassified as benign and becomes an adversarial example. Although the figure is depicted in two dimensions for simplicity, the actual manipulation space includes multiple features such as flow duration, packet count, and packet size (see Table~\ref{tab:feature_modifiable}). \textcolor{black}{In Figure~\ref{fig:advtransfo}, the labels \textit{Class A, Class B, Class C, and Class D} correspond to different attack types, each representing a distinct category of malicious traffic that is gradually transformed into an adversarial example.} The adversarial perturbation process is formalized in Equation~\ref{eq:moyen}:
\begin{equation}
\begin{aligned}
\epsilon_t \;=\; c\,t\,\bigl\lVert M_b - x_0 \bigr\rVert, 
\qquad
x_{t}^{\mathrm{adv}}(f) \;=\; 
\mathrm{Project}\Bigl[x_0(f)\;+\;\epsilon_t \;\cdot\;\mathrm{sign}\bigl(M_b(f)\;-\;x_0(f)\bigr)\Bigr].
\end{aligned}
\label{eq:moyen}
\end{equation}

Where:
\begin{itemize}
    \item[$\bullet$] \textbf{$f$:} one of the three modifiable network features (i.e., \emph{Dur}, \emph{Spkts}, \emph{Sbytes}).
    \item[$\bullet$] \textbf{$x_{t}^{\mathrm{adv}}(f)$:} the value of the adversarial instance at iteration $t$ for feature $f$.
    \item[$\bullet$] \textbf{$x_{0}(f)$:} the initial value of feature $f$ in the malicious instance before any perturbation.
    \item[$\bullet$] \textbf{$M_b(f)$:} the mean value of feature $f$ across all benign instances.
    \item[$\bullet$] \textbf{$\epsilon_t$:} the step size at iteration $t$, defined as 
    \[
        \epsilon_t \;=\; c\,t\,\bigl\lVert M_b - x_0 \bigr\rVert.
    \]
    Here, $\lVert M_b - x_0 \rVert$ is the Euclidean distance between the initial malicious instance $x_0$ and the benign centroid $M_b$, and $c$ is a small constant tuning the perturbation rate.
    \item[$\bullet$] \textbf{$\mathrm{sign}(\cdot)$:} a function that returns $+1$ if its argument is positive and $-1$ if its argument is negative, thereby indicating the direction in which the feature should move.
    \item[$\bullet$] \textbf{$\mathrm{Project}(\cdot)$:} a projection function that ensures each updated feature value remains within valid syntactic and semantic bounds. In practice, any coordinate that falls outside its allowable range is clipped or mapped to the nearest acceptable value.
\end{itemize}

\textcolor{black}{
\textbf{Role of the \textit{Project()} Function:} In Equation~\ref{eq:moyen}, after adding a scaled directional perturbation to the previous feature value, the \textit{Project()} function maps the result back into a valid range. This step is crucial for two reasons:
\begin{enumerate}
    \item \textbf{Syntactic Constraints:} It ensures that each network feature (e.g., duration must be non-negative and within realistic limits) remains in its allowable format and range.
    \item \textbf{Semantic Constraints:} It preserves the logical relationships among features. For example, if packet count and byte size are interdependent, the projected values maintain consistency between them.
\end{enumerate}
Additionally, the \textit{Project()} function in Equation~\ref{eq:moyen} enforces both a lower and an upper bound on the perturbed feature values. If a modification exceeds the permissible range - defined by these syntactic and semantic constraints - the function clips or maps the value to the nearest acceptable boundary. This mechanism ensures that adversarial instances remain both valid and realistic within the expected constraints of network traffic.
By enforcing these constraints, the \textit{Project()} function guarantees that any adversarial instance remains plausible, thus enhancing the stealth of the attack.
}

\begin{figure}
    \centering
    \includegraphics[width=1.0\linewidth]{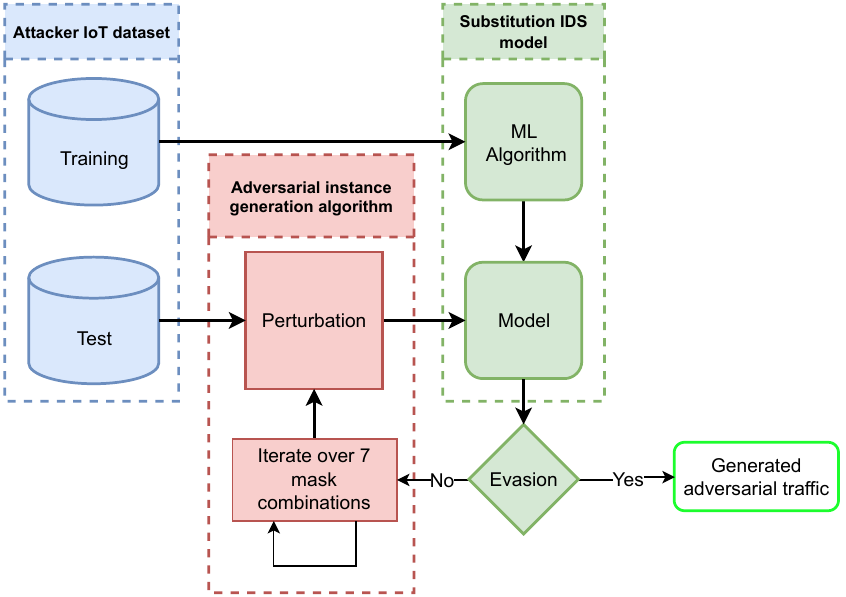}
    \caption{Adversarial instance generation process. The attacker’s IoT dataset is split to build a substitution model that approximates the defender’s IDS. Iterative perturbations (and mask switching) are then applied to malicious flows until successful evasion is achieved, producing adversarial traffic ready to bypass the target system.}
    \label{fig:adversarial_generation_process}
\end{figure}

\textcolor{black}{Figure~\ref{fig:adversarial_generation_process} illustrates the overall workflow of our adversarial instance generation approach. Starting from the attacker IoT dataset, we split the data into training and test subsets. The training portion is used to build a substitution model that emulates the defender’s IDS. Our adversarial generation algorithm then iteratively applies perturbations (and, if needed, switches masks) to malicious flows, ultimately producing adversarial traffic capable of evading the IDS. Once the algorithm confirms successful evasion, the generated adversarial samples can be launched against the target system.}

\begin{algorithm}[h!]
\caption{Crafting adversarial examples using the D2TC method}
\begin{algorithmic}[1]
\Procedure{CraftAdvEx}{x} \Comment{where $x$ is a malicious flow}
\State $t \gets 1$
\Repeat
    \State $m \gets Benign\_dist(x(f),M_b(f))$
    \For{$\text{mask} \gets \text{mask}_1,\dots,\text{mask}_7$}
        \State $\epsilon \gets \text{Sign}(M_b(f)- x(f)) \cdot (c \cdot t) \cdot m \cdot \text{mask}$
        \State $x_{adv} \gets x + \epsilon$
        \State $x_{adv} \gets Project(x_{adv})$
        \If{$predict(x_{adv}) = \text{benign}$}
            \State \textbf{return} $x_{adv}$
        \EndIf
    \EndFor
    \State $t \gets t + 1$
\Until{$t = T_{max}$} \Comment{$T_{max}$ is set to 10 in our experiments}
\EndProcedure
\end{algorithmic}
\label{alg:craftadvexalgo}
\end{algorithm}

The iterative process is detailed in Algorithm~\ref{alg:craftadvexalgo}. Starting with a small perturbation, the algorithm gradually increases the perturbation until the adversarial instance is misclassified as benign. The constant $c$ controls the perturbation rate, ensuring that only minimal changes are applied initially to reduce detection likelihood.

\textcolor{black}{To further minimize unnecessary modifications, our method employs a binary mask approach. Each mask is represented as a binary string, where each digit corresponds to a specific modifiable network feature (e.g., \emph{Dur}, \emph{Spkts}, \emph{Sbytes}). A value of 1 indicates that the feature is perturbed, while 0 means it remains unchanged. As summarized in Table~\ref{tab:combinaisons}, each mask uniquely defines a combination of feature perturbations. For instance, mask 1 (binary \textbf{001}) perturbs only the \emph{Dur} feature, mask 3 (binary \textbf{100}) perturbs only the \emph{Sbytes} feature, and mask 7 (binary \textbf{111}) indicates that all three features are perturbed simultaneously.\\
Within Algorithm~\ref{alg:craftadvexalgo}, the loop over the masks (mask1,..., mask7) iterates over this predefined set of binary masks. This loop allows the algorithm to explore various combinations of feature perturbations sequentially, thereby determining the minimal set of modifications required to fool the IDS while preserving the semantic integrity of the network traffic.}

\begin{table}[h!]
    \centering
    \caption{ Binary masks for modifiable network features (\emph{Dur}, \emph{Spkts}, \emph{Sbytes}), indicating which combination of features is perturbed for each mask ID}
    \label{tab:combinaisons}
    \begin{tabularx}{0.8\textwidth}{>{\centering\arraybackslash}m{2cm} >{\centering\arraybackslash}m{2cm} X}
        \toprule
        \textbf{Mask ID} & \textbf{Mask Value} & \textbf{Manipulated Features} \\
        \midrule
        1  & 001 & $Dur$ \\
        2  & 010 & $Spkts$ \\
        3  & 100 & $Sbytes$ \\
        4  & 011 & $Dur$, $Spkts$ \\
        5  & 101 & $Dur$, $Sbytes$ \\
        6  & 110 & $Sbytes$, $Spkts$ \\
        7  & 111 & $Dur$, $Sbytes$, $Spkts$ \\
        \bottomrule
    \end{tabularx}
\end{table}

Our adversarial algorithm incorporates two objective functions: one that minimizes the magnitude of change in each feature (via Equation~\ref{eq:moyen}), and another that minimizes the number of altered features using the \textit{Mask Method}. This dual-objective strategy leverages $\ell_2$-norm regularization to control overall perturbation magnitude, while $\ell_0$-norm regularization limits the number of modified features.

\color{black}

\paragraph{How D2TC Minimizes Feature Perturbations}
\label{Sec:MathProof}
D2TC aims to minimize the perturbation required to evade detection by incrementally guiding a malicious instance $x_0$ toward the benign class centroid 
\[
M_b = \frac{1}{|B|}\sum_{x\in B} x,
\]
where $B$ is the set of benign training samples. The direction of perturbation is given by the unit vector 
\[
d = \frac{M_b - x_0}{\|M_b - x_0\|},
\]
and the step size increases linearly with each iteration according to  
\[
\epsilon_t = c\,t\,\bigl\|M_b - x_0\bigr\|,
\]
where $c$ is a small scaling factor. Each update is computed as 
\[
x_t = \mathrm{Project}\bigl(x_0 + \epsilon_t\,d\bigr),
\]
where \texttt{Project()} ensures that the resulting instance remains valid by enforcing syntactic and semantic constraints (i.e., bounding features within permissible ranges).

To illustrate the rationale behind this design, let's consider the case without projection. In this idealized scenario, the cumulative perturbation after $t$ steps is
\[
x_t - x_0 = \sum_{i=1}^t \epsilon_i\,d = \left(c \cdot \frac{t(t+1)}{2}\right)\,d,
\]
which remains collinear with $M_b - x_0$. Hence, the total $\ell_2$-norm perturbation is
\[
\|x_t - x_0\| = c\,\frac{t(t+1)}{2}\,\|M_b - x_0\|.
\]
This implies that D2TC follows a straight-line trajectory toward the benign region, which is theoretically the most efficient path in Euclidean space. While projection may alter this trajectory slightly to maintain plausibility, the core idea remains: D2TC heuristically follows the direction that yields the least change in feature space under the constraint of step-wise misclassification.

Importantly, we emphasize that D2TC does not guarantee globally minimal perturbations. Rather, it seeks to minimize perturbations by applying incremental updates in a direction that is empirically likely to induce misclassification while preserving realistic traffic semantics. This balance allows D2TC to generate effective, valid adversarial examples with minimal modification to the original malicious flows.

\color{black}
\section{Strengthening adversarial robustness}
\label{sec:def}

Evasion attackers attempt to deceive network intrusion detection systems by introducing precisely crafted perturbations into the network traffic, aiming to maximize classification errors. These alterations in network traffic, often subtle and hard to detect, can be identified by adversarial detectors specifically trained to recognize such malicious modifications. Adversarial detection is thus a crucial first line of defense, focused on identifying whether incoming traffic has been manipulated with adversarial intent.

Building on the foundation of adversarial detection, our proposed defense strategy enhances the resilience of the NIDS by combining detection techniques with an ensemble method known as Bootstrap Aggregating (Bagging)~\cite{breiman1996bagging}. The primary goal of this defense is to filter out adversarial examples before they reach the NIDS, thereby preventing potential attacks from disrupting network security. By employing multiple adversarial detectors working in tandem during the inference phase, our approach increases the likelihood of identifying and eliminating adversarial perturbations, effectively reducing the risk of evasion attacks.

The key difference between standard adversarial detection and our proposed defense strategy lies in their scope and effectiveness. While adversarial detection focuses on recognizing and labeling adversarial inputs, our strategy goes further by integrating these detection methods into an ensemble framework. This approach not only detects adversarial inputs but also mitigates their impact by filtering them out before they can cause harm. In essence, our defense strategy strengthens the overall robustness of the NIDS, ensuring a more comprehensive and reliable protection against adversarial attacks.

\subsection{Data partitioning}
\label{sec:DataPartitioning}

\begin{figure}
    \centering
    \includegraphics[width=0.9\linewidth]{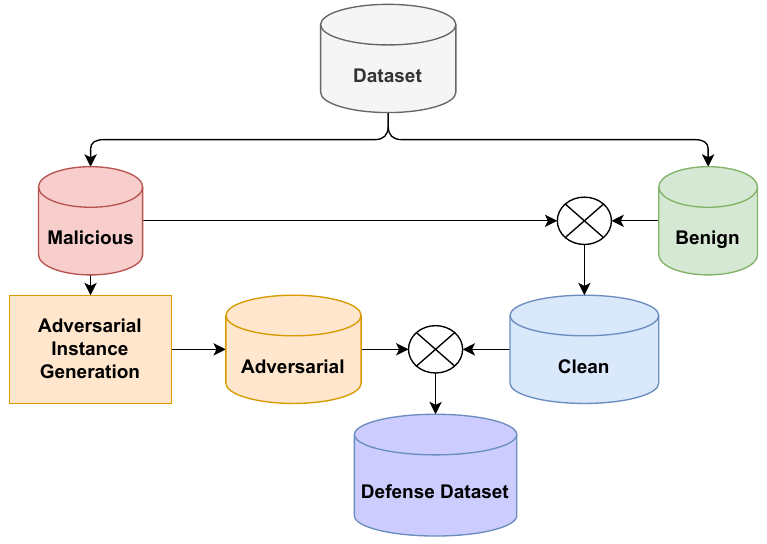}
    \caption{Workflow for creating the defense dataset. The original dataset is split into benign and malicious flows, with malicious instances undergoing adversarial instance generation (D2TC) to produce adversarial samples. These adversarial and clean subsets are then combined to form a comprehensive defense dataset.}
    \label{fig:Def_Data_Gen}
\end{figure}

\textcolor{black}{As illustrated in Figure~\ref{fig:Def_Data_Gen}, we begin by separating the original dataset into two main categories: \emph{benign} and \emph{malicious}. The malicious portion undergoes adversarial instance generation (via the D2TC method), resulting in a set of \emph{adversarial} flows, while the \emph{clean} subset is composed of unmodified benign and malicious instances. These two subsets are then combined to form a specialized \emph{defense dataset} that includes both adversarial and clean traffic. This comprehensive dataset ensures that our defense approach can learn to distinguish subtle perturbations introduced by adversarial examples from legitimate network behavior.}

\begin{figure}[h]
\centering
\includegraphics[width=1.0\columnwidth]{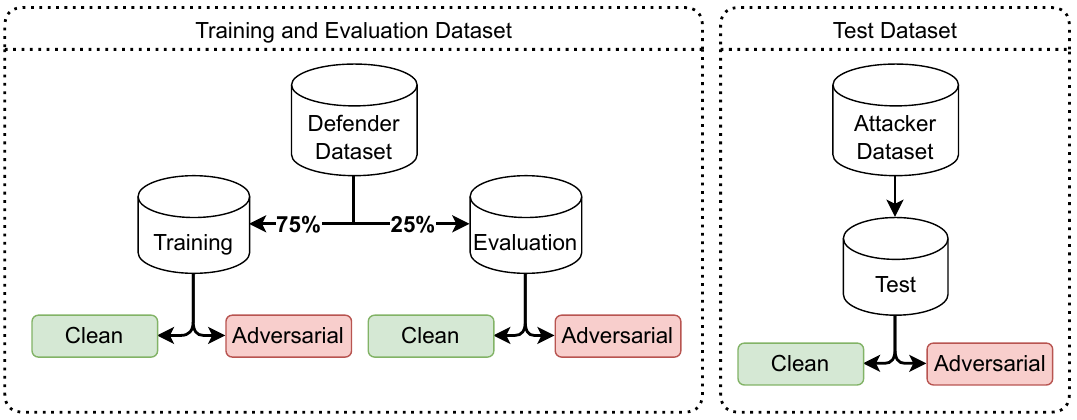}
\caption{Partitioning of the defense dataset into training/evaluation (derived from the defender’s data) and testing (sourced from the attacker’s data)}
\label{fig:partitioning_datasets}
\end{figure}

\textcolor{black}{After that, as depicted in Figure~\ref{fig:partitioning_datasets}, we partition the defense dataset into training and evaluation sets to train our detectors in a realistic setting. Specifically, the training set is sourced from the defender’s portion of the dataset, providing real-world representations of benign and adversarial instances. Meanwhile, an entirely separate test set, derived from the attacker’s dataset, allows us to assess the generalization capability of our defense strategy on previously unseen data. This two-tier partitioning scheme helps prevent overfitting and ensures that our results accurately reflect performance under genuine adversarial conditions.}

\subsection{Defense Process}
\label{sec:DefenseProcess}

\begin{figure}[!hbt]
\centering
\includegraphics[width=0.85\columnwidth]{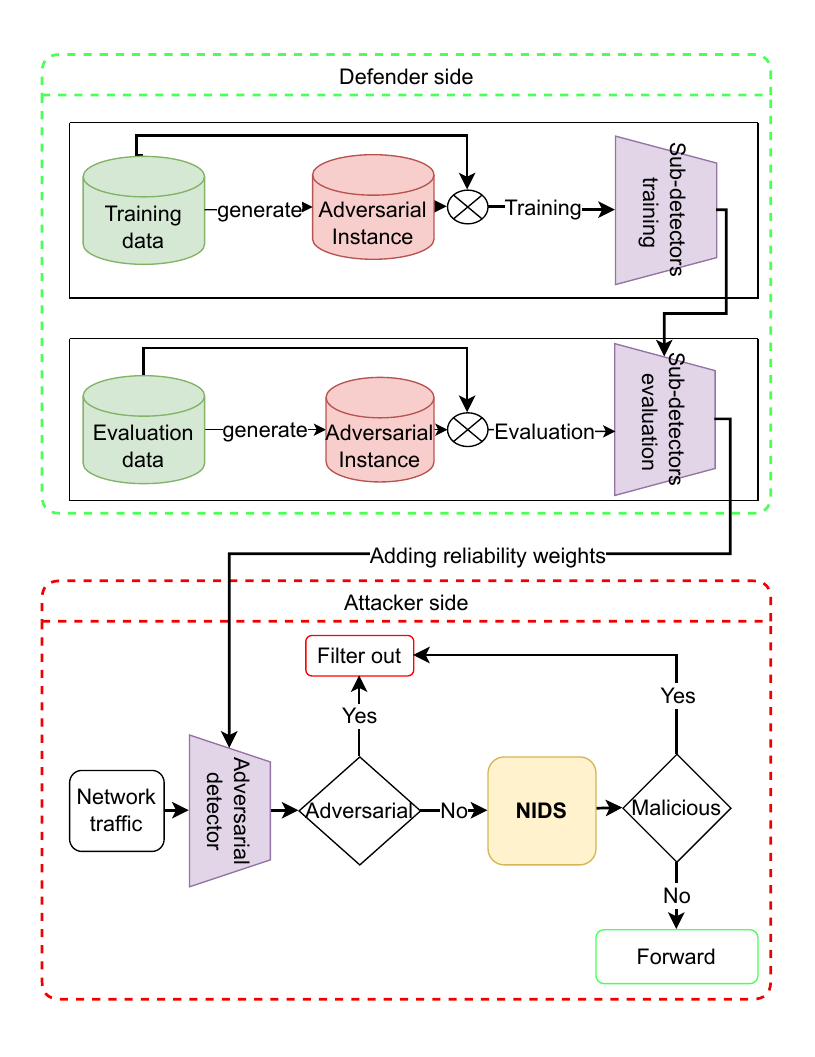}
\caption{Overview of the proposed defense process. Sub-detectors are trained and evaluated in parallel on labeled clean and adversarial data, with each sub-detector assigned a reliability weight based on its detection performance. The final, ensemble-based decision filter is placed before the NIDS to intercept and mitigate adversarial traffic}
\label{fig:defprocess}
\end{figure}

The proposed defense method, as depicted in Figure~\ref{fig:defprocess}, follows a reactive strategy consisting of several steps:

\begin{enumerate}
    \item \textbf{Generating a labeled dataset}: A dataset is created containing both clean and adversarial instances. The dataset generation algorithm must produce a representative sample sufficient for effectively training the sub-detectors.
    
    \item \textbf{Training sub-detectors}: Thirteen (13) sub-detectors are trained in parallel, each specializing in a single feature (the total number of features is thirteen, as detailed in Tables~\ref{tab:feature_modifiable}, \ref{tab:feature_dependent}, and \ref{tab:feature_non_modifiable}). This specialization allows each sub-detector to learn the distinctive patterns of clean versus adversarial instances based on its assigned feature.
    
    \color{black}
    \item \textbf{Evaluating sub-detectors}: The performance of each sub-detector is evaluated using a separate evaluation dataset. The detection rate (Recall) is used as the primary metric to assess each sub-detector's ability to correctly identify adversarial instances.
    
    \item \textbf{Determining sub-detector reliability}: Based on the evaluation results, the detection rate values for each sub-detector are normalized to compute reliability weights. These weights indicate the trustworthiness of each sub-detector's predictions and determine the influence that each sub-detector will have on the final decision. An adaptive weighting mechanism is employed to update these weights based on performance on an independent evaluation dataset, mitigating potential inductive bias in assigning importance to individual features.
    
    \color{black}
    
    \item \textbf{Balancing sub-detector decisions}: The reliability weights are used to weight the decisions of each sub-detector. These weighted outputs are then aggregated using ensemble fusion techniques (e.g., Bayesian fusion and Dempster-Shafer combination) to produce a unified and balanced final decision.
    
    \item \textbf{Deploying the combined detector before the NIDS}: The combined detector, which integrates the outputs of all sub-detectors, is positioned ahead of the NIDS. This placement intercepts and filters adversarial traffic before it reaches the IDS, thereby enhancing overall security and system performance.
\end{enumerate}

\subsection{Defense architecture}
\label{fusionrules}

\begin{figure*}[h!]
\centering
\includegraphics[width=\textwidth]{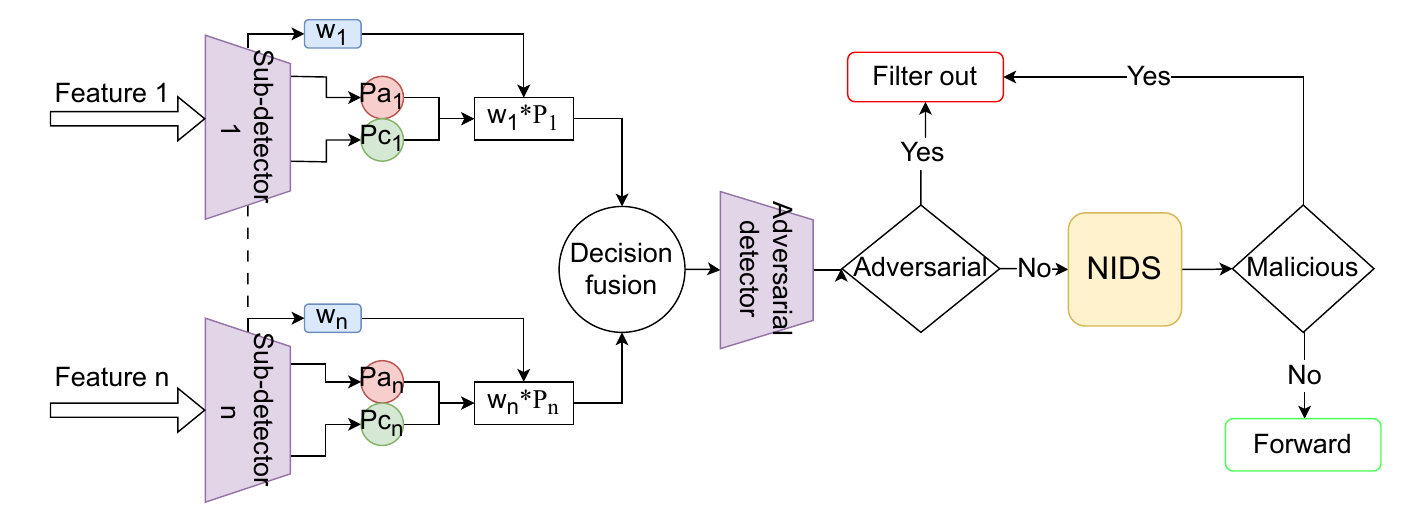}
\caption{Overview of the proposed adversarial detector architecture. Each incoming instance is split into feature-specific sub-detectors, whose decisions are weighted by reliability scores and fused to determine whether the traffic is benign or adversarial}
\label{architecture of defense}
\end{figure*}

As shown in Figure \ref{architecture of defense}, The defense architecture processes each instance by splitting it into 13 feature values, each of which is sent to a corresponding sub-detector for analysis. Each sub-detector computes a decision value, which is then multiplied by its reliability weight. The weighted decisions are combined using a fusion function to determine whether the instance is benign or adversarial.

In our defense strategy, two fusion rules are evaluated:

\textbf{Bayesian fusion}: This rule, detailed in Equation \ref{eq: bayésienne}, combines the predictions from all sub-detectors by integrating their prediction uncertainties and applying reliability weights to each sub-detector's output. 

\begin{equation}
\label{eq: bayésienne}
\begin{aligned}
P_a &= \frac{\sum_{i=1}^{13} P_{ai} \cdot w_i}{\sum_{i=1}^{13} P_{ai} \cdot w_i + \sum_{i=1}^{13} P_{ci} \cdot w_i}, \\
P_c &= \frac{\sum_{i=1}^{13} P_{ci} \cdot w_i}{\sum_{i=1}^{13} P_{ai} \cdot w_i + \sum_{i=1}^{13} P_{ci} \cdot w_i}
\end{aligned}
\end{equation}

Here, \(P_a\) and \(P_c\) represent the probabilities of the instance being adversarial or clean, respectively. The index \(i\) refers to the \(i\)th sub-detector, and \(w_i\) is the corresponding reliability weight for that sub-detector.

\textbf{Dempster-Shafer combination}: This rule integrates evidence from multiple sources (i.e., sub-detectors) to establish a belief level for each classification decision.

Let $\Omega = \{\omega_1, \ldots, \omega_K\}$, and $P(\Omega) = \{A_1, \ldots, A_Q\}$ denote its power set, where $Q = 2^K$. A mass function $M: P(\Omega) \rightarrow [0,1]$ is a basic belief assignment if $M(\emptyset) = 0$ and $\sum_{A \in P(\Omega)} M(A) = 1$.

When two bases $M_1$ and $M_2$ represent evidence (i.e., decisions from two detectors), the Dempster-Shafer fusion rule combines them, resulting in $M = M_1 \oplus M_2$ as defined in Eq~\ref{eq:DS}. 

\begin{equation}
\label{eq:DS}
    M(A) = (M_1 \oplus M_2)(A) \propto \sum_{B_1 \cap B_2 = A} M_1(B_1) M_2(B_2)
\end{equation}

\section{Evaluation}
\label{sec:results}

In this section, we present the findings from various experiments. We start by discussing the performance of the initial attacker and defender models, evaluating different metrics such as \textbf{precision, recall, and F1 score}. These results are based on models trained using the Ton-IoT and Bot-IoT datasets, as detailed in the following subsections.

Then, we explore the performance of the models in adversarial scenarios. We conduct a preliminary study on models trained with the Ton-IoT dataset to analyze transferability between different models. Additionally, we examine the time required for each attack and compare the perturbation levels between the original malicious instances and the adversarial instances. Furthermore, we provide a comprehensive comparison of the proposed adversarial generation algorithm's performance across the various attacks in each dataset.

Finally, we present the results of the proposed defense mechanism against the adversarial instances generated by the previously discussed evasion attack algorithm.

\textcolor{black}{To comprehensively evaluate the performance of our adversarial attack and defense mechanisms, we employ several well-established metrics, namely precision, recall, and F1-score. \textbf{Precision} is defined as the ratio of true positive detections to the total number of instances classified as positive. In the context of IDS, this metric measures the proportion of correctly identified attack instances among all instances flagged as malicious, reflecting the system's ability to minimize false alarms. \textbf{Recall} (or detection rate) is the ratio of true positive detections to the actual number of positive instances, indicating how effectively the system captures actual attacks. The \textbf{F1-score}, the harmonic mean of precision and recall, provides a balanced evaluation particularly useful in scenarios with class imbalance, which is common in network intrusion detection where attack instances are rare compared to benign traffic.\\
We selected these metrics because they offer a nuanced view of IDS performance beyond simple accuracy. In adversarial settings - where both false positives and false negatives can have severe consequences - precision, recall, and F1-score allow for a more detailed assessment of how well the IDS discriminates between malicious and benign traffic under adversarial conditions.}

\subsection{Datasets}
The ToN-IoT~\cite{moustafa2021new} and BoT-IoT~\cite{koronIoTis2019towards} datasets were used to ensure reproducibility of the experiments and allow for comparable results across multiple datasets, as well as to cover a wider range of attacks.
\begin{itemize}

    \item  \textbf{ToN-IoT dataset}: provide a variety of data sources obtained from IoT and IIoT telemetry datasets, operating system datasets and network traffic datasets. They were collected from a realistic, large-scale network created at UNSW Canberra's \textit{Cyber Range} and IoT Labs. The network included a new Industry 4.0 testbed with multiple virtual machines and hosts running Windows, Linux and Kali operating systems. Data sets were collected in parallel to capture normal events and cyber attacks, including DoS, DDoS and ransomware attacks against web applications, IoT gateways and computer systems in the IoT/IIoT network.
    \item  \textbf{BoT-IoT dataset}: is a collection of normal and botnet traffic created in a realistic network environment at the \textit{Cyber-Range} Lab at UNSW Canberra. The dataset provides source files in a variety of formats, including pcap, argus, and csv files, categorized by attack subcategories for easy labeling. The pcap files account for over 69 GB with 72 million records, while the extracted csv stream traffic accounts for 16.7 GB. Attacks included are DDoS, DoS, OS and Service Scan, Keylogging, and Data exfiltration, organized according to the protocol used. To simplify handling, a 5\% subset of the dataset was extracted, comprising approximately 3 million records spread across four files with a total size of 1.07 GB.

\end{itemize}

\color{black}

\subsection{Training and Validation Process}
\label{sec:TrainValProc}
To simulate a realistic black-box attack scenario, we divide each dataset (BoT-IoT and ToN-IoT) into two equal parts: one for the defender and one for the attacker. Specifically, each dataset portion contains over 24,000 instances for BoT-IoT and over 40,000 instances for ToN-IoT, ensuring a balanced mix of benign and malicious traffic. This division allows the attacker to build substitute  models using a dataset similar to, but not identical with, the defender’s. 

\begin{figure}[h!]
\centering

\includegraphics[width=\columnwidth]{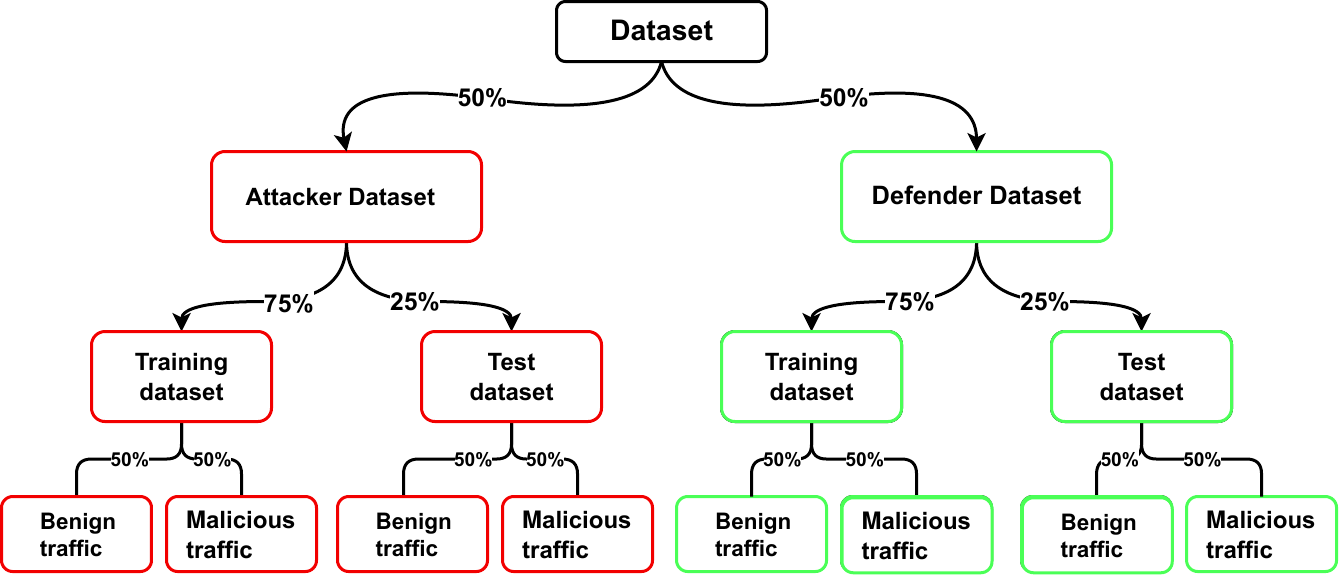}
\caption{Partitioning of the BoT-IoT and ToN-IoT datasets into attacker and defender subsets, each further split into training (75\%) and test (25\%) sets}
\label{fig:partitioning}
\end{figure}

As depicted in Figure~\ref{fig:partitioning}, each portion of the dataset is further split into training (75\%) and test (25\%) subsets. This step ensures that both the defender and attacker have distinct training and test sets, preventing data leakage and more accurately reflecting real-world conditions where attackers do not have access to the defender’s exact training data. The 75\%–25\% split is chosen to provide sufficient data for model training while reserving enough data for robust performance evaluation.

\paragraph{Training Procedure}

\begin{enumerate}
    \item \textbf{Data Preparation:} We apply our preprocessing pipeline (Section~\ref{sec:preprocessing}), which includes data cleaning, normalization, and transformation (e.g., one-hot encoding for categorical features).
    \item \textbf{Model Selection:} We then choose our machine learning algorithms (KNN, RF, DT, and XGBoost) based on their suitability for tabular data and computational efficiency in IoT settings (Section~\ref{sec:mlrationale}).
    \item \textbf{Hyperparameter Tuning:} We perform a combined grid search and random search on the training subset to identify optimal hyperparameters for each model. This includes varying parameters like the number of neighbors (KNN), the number of estimators (RF, XGBoost), and splitting criteria (DT).
    \item \textbf{Training Phase:} Using the best-found hyperparameters, we train each model on the 75\% training subset. This ensures the models learn from a representative sample of benign and malicious traffic.
    \item \textbf{Validation and Attacker/Defender Split:} While the defender trains and validates the NIDS using the first portion of the dataset, the attacker trains substitute  models on the second portion. This separation mimics a realistic adversarial setting where the attacker has “similar but not identical” data.
\end{enumerate}

\paragraph{Model Validation}
Once training is complete, each model is evaluated on its respective 25\% test subset. This test set remains unseen during training, ensuring an unbiased estimate of the model’s generalization performance. We track key metrics (e.g., precision, recall, F1-score) to measure the IDS’s ability to detect malicious traffic accurately. Additionally, for adversarial evaluations, the attacker uses their trained substitute  models to generate adversarial samples, which are then tested against the defender’s NIDS to assess robustness. This two-tier validation scheme provides a comprehensive view of both standard performance and adversarial resilience.

Overall, this step-by-step process ensures that both defender and attacker models are trained and tested under conditions that closely resemble real-world IoT network scenarios, facilitating a fair and robust assessment of the proposed methods.

\color{black}

\color{black}

\subsection{Preprocessing}
\label{sec:preprocessing}
Before introducing the data into our models, we perform several preprocessing steps to ensure the accuracy and suitability of the data for analysis. These steps include data cleaning, normalization, and data transformation.

\subsubsection*{Data Cleaning}
Data cleaning involves identifying and correcting errors in the raw dataset to improve its quality and reliability. Our data cleaning process includes:
\begin{enumerate}
    \item \textbf{Removing Duplicates:} We identify duplicate records by comparing feature values across the dataset and remove redundant entries to avoid bias in the analysis.
    \item \textbf{Correcting Typographical Errors:} We both manually and automatically examine the data to correct typos. For example, we verify that the values of certain features conform to known relationships or formulas (e.g., the expected correlation between packet counts and byte counts).

\end{enumerate}

\subsubsection*{Normalization}
Normalization transforms raw data into a uniform scale, which is crucial for many machine learning models. In our work, we employ:
\begin{itemize}
    \item \textbf{Min-max Normalization:} This method scales each feature to a [0, 1] range using the formula:
    \begin{equation}
    \label{eq:minmaxscale}
    X_{\text{scaled}} = \frac{X - \min(X)}{\max(X) - \min(X)},
    \end{equation}
    where $X$ is the original value and $X_{\text{scaled}}$ is the normalized value. We use this method because it is simple and effective at improving model performance.
    \item \textbf{Z-score Normalization:} Although mentioned as an alternative, we did not adopt this method, as our experiments showed that min-max normalization provided better performance for our dataset.
\end{itemize}

\subsubsection*{Data Transformation}
Data transformation converts data from its original format into a structure more suitable for analysis and model training. Our transformation process includes:

    \textbf{Categorical Encoding:} Categorical features such as "Proto", "Sport", and "Dport" are converted into numerical representations via one-hot encoding. For the features "Sport" and "Dport", which are originally numeric, we first recategorize them into discrete ranges before applying one-hot encoding. This step is crucial for capturing the underlying patterns associated with these features.

\begin{table}[h!]
    \centering
    \caption{Port range categorization for the \texttt{Sport} and \texttt{Dport} features prior to one-hot encoding}
    \label{tab:port_categorization}
    \begin{tabularx}{0.8\textwidth}{C C}
        \toprule
        \textbf{Port Range} & \textbf{Category} \\
        \midrule
        0 to 1023 & Well-Known Ports \\
        1024 to 49151 & Registered Ports \\
        49152 to 65535 & Dynamic/Private Ports \\
        \bottomrule
    \end{tabularx}
\end{table}

\color{black}

\subsection{Machine Learning-based NIDS}
\label{sec:mlrationale}
\color{black}
Unlike deep learning (DL) approaches, ML models offer advantages in terms of computational efficiency and performance on tabular data, making them well-suited for resource-constrained IoT environments~\cite{apruzzese2020hardening, elsayed2022litelstm}.

In this work, we employ four distinct machine learning algorithms - K-Nearest Neighbors (KNN), Random Forest (RF), Decision Trees (DT), and XGBoost - to develop IDS models for both the defender and the attacker. KNN is chosen for its simplicity and interpretability, which are beneficial when working with tabular network data. RF is selected because it is robust to overfitting and offers reliable feature importance insights, while DTs provide clear decision logic that aids in understanding model behavior. XGBoost is incorporated due to its high predictive performance and efficiency, especially when handling large datasets. By utilizing this diverse set of algorithms, we capture a wide range of sensitivities to adversarial perturbations, enabling a more comprehensive evaluation of model transferability and robustness.

Hyperparameter selection and tuning were performed using a combination of grid search and random search methods. For instance, we varied the number of neighbors for KNN, adjusted the number of estimators for RF and XGBoost, and experimented with different splitting criteria for DTs. To prevent the attacker from exploiting identical model configurations, we ensured that each model was trained with distinct random hyperparameters. This approach not only guarantees fair comparisons between models but also simulates realistic scenarios where adversaries have limited insight into the defender’s settings.
\color{black}

\begin{figure*}[h!]
\centering
\includegraphics[width=\columnwidth]{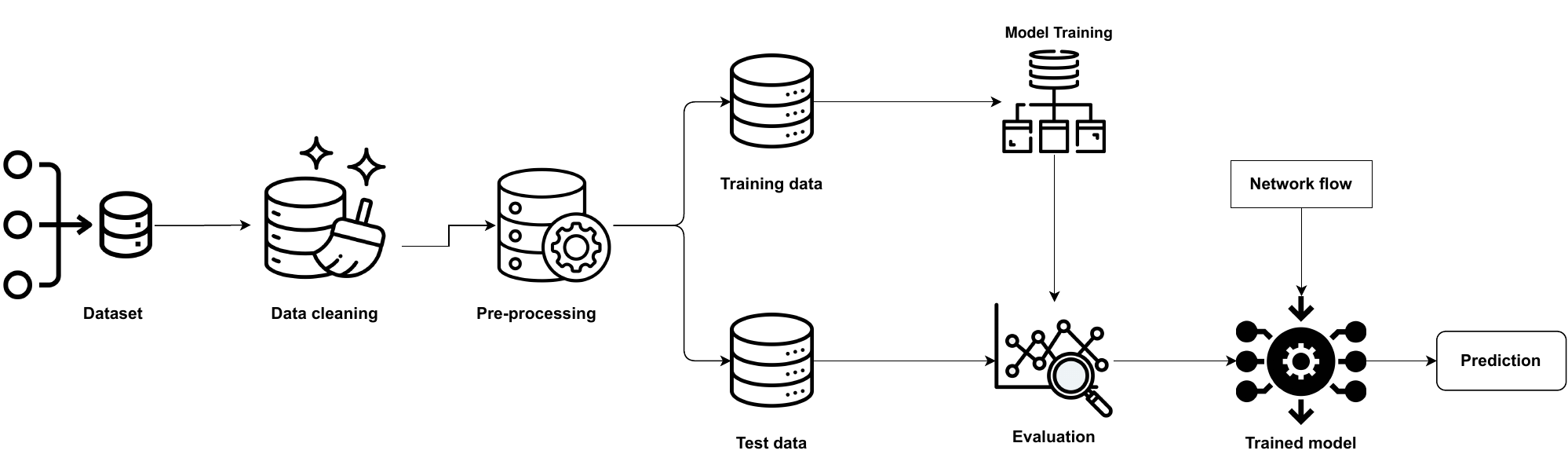}
\caption{Machine Learning pipeline for both attacker and defender}
\label{fig:mlpipeline}
\end{figure*}

The training and testing processes for all models follow the same pipeline, as illustrated in Figure~\ref{fig:mlpipeline}. This standardized process further supports a balanced and transparent evaluation of transferability and performance across various models in the presence of adversarial attacks.

\begin{figure}[h!]
\centering
\includegraphics[width=\columnwidth]{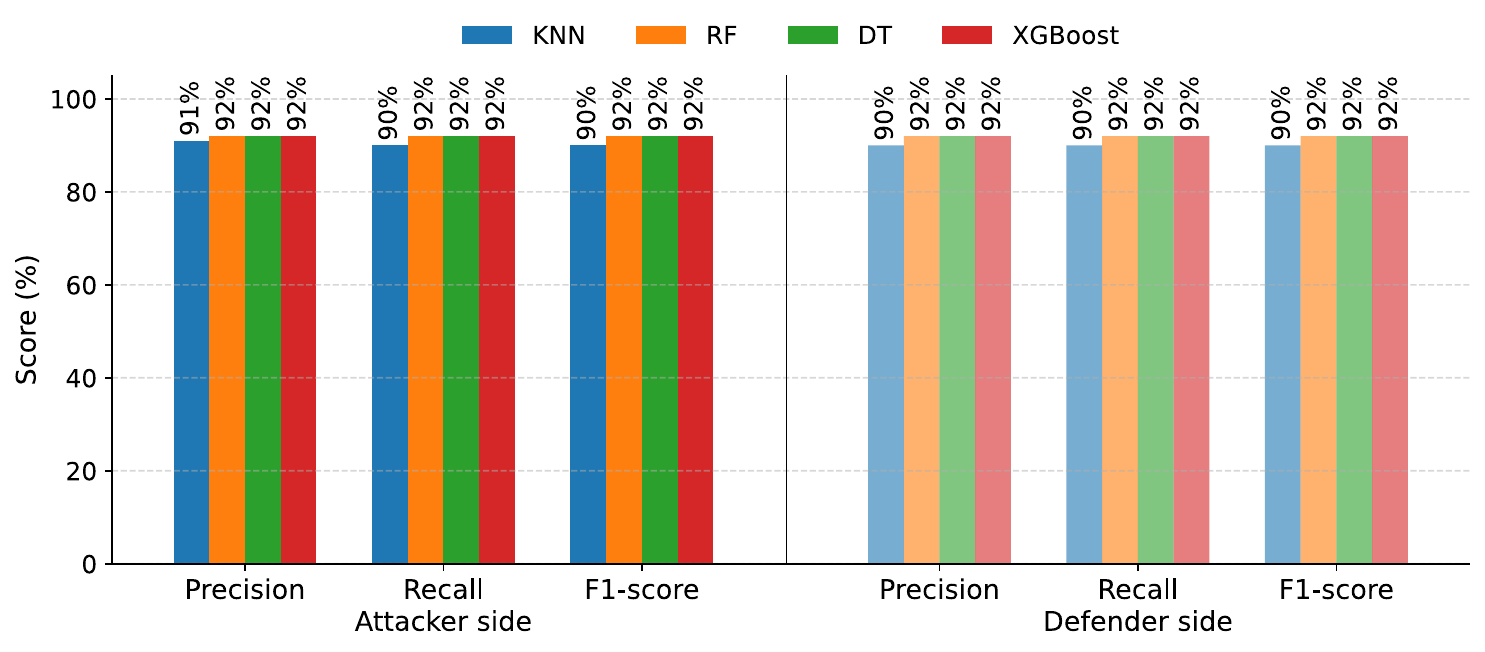}
\caption{Initial performance of the trained models (KNN, RF, DT, XGBoost) on both the attacker and defender sides for the Bot-IoT dataset. All models exhibit high metric scores (precision, recall, F1-score), with minimal variations across different hyperparameters.}
\label{model-def-att}
\end{figure}

\begin{figure}[h!]
\centering
\includegraphics[width=\columnwidth]{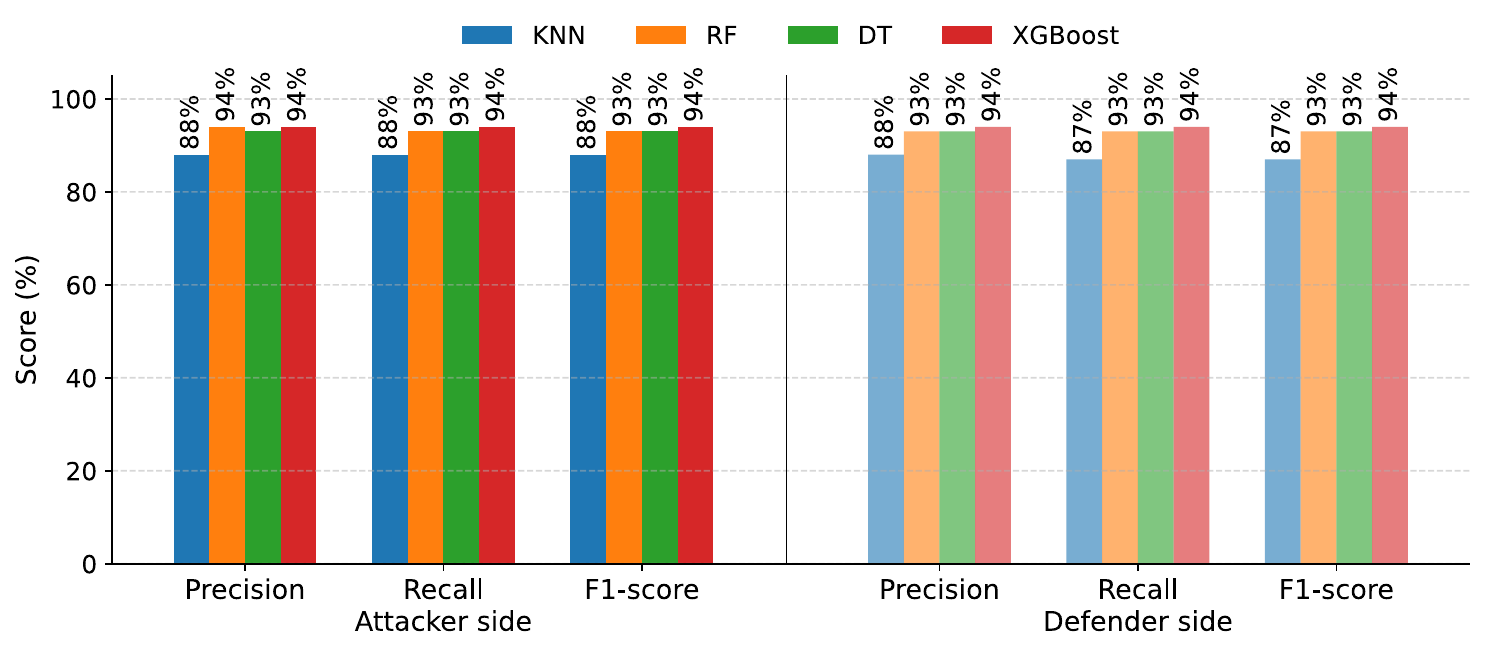}
\caption{Initial performance of the trained models (KNN, RF, DT, XGBoost) on both the attacker and defender sides for the Ton-IoT dataset. Similar to the Bot-IoT results, each model achieves consistently high metric scores, demonstrating negligible differences despite variations in hyperparameters.}
    \label{model-def-att-TON}
\end{figure}

The initial performance of the KNN, RF, DT, and XGBoost models on both the attacker and defender sides is shown in Figure~\ref{model-def-att} for the Bot-IoT dataset and in Figure~\ref{model-def-att-TON} for the Ton-IoT dataset. In both cases, each model achieves consistently high metric scores (precision, recall, F1-score), exceeding 90\%. Moreover, when comparing the defender’s models to the attacker’s substitute models, the observed variations are negligible - around 1\% at most - indicating that differences in hyperparameters have minimal impact on overall performance.

\subsection{Impact of adversarial perturbations}
\label{sec:ImpAdvPert}

\color{black}
In this subsection, we evaluate the performance of our proposed adversarial attack against the defender IDSs. Specifically, we explore the transferability of adversarial instances generated by the attacker's models to the defender's models, and analyze results from various perspectives, including the number of masks used, levels of perturbation, and the time required for generation.

\begin{figure}[h]
  \centering
\includegraphics[width=\columnwidth]{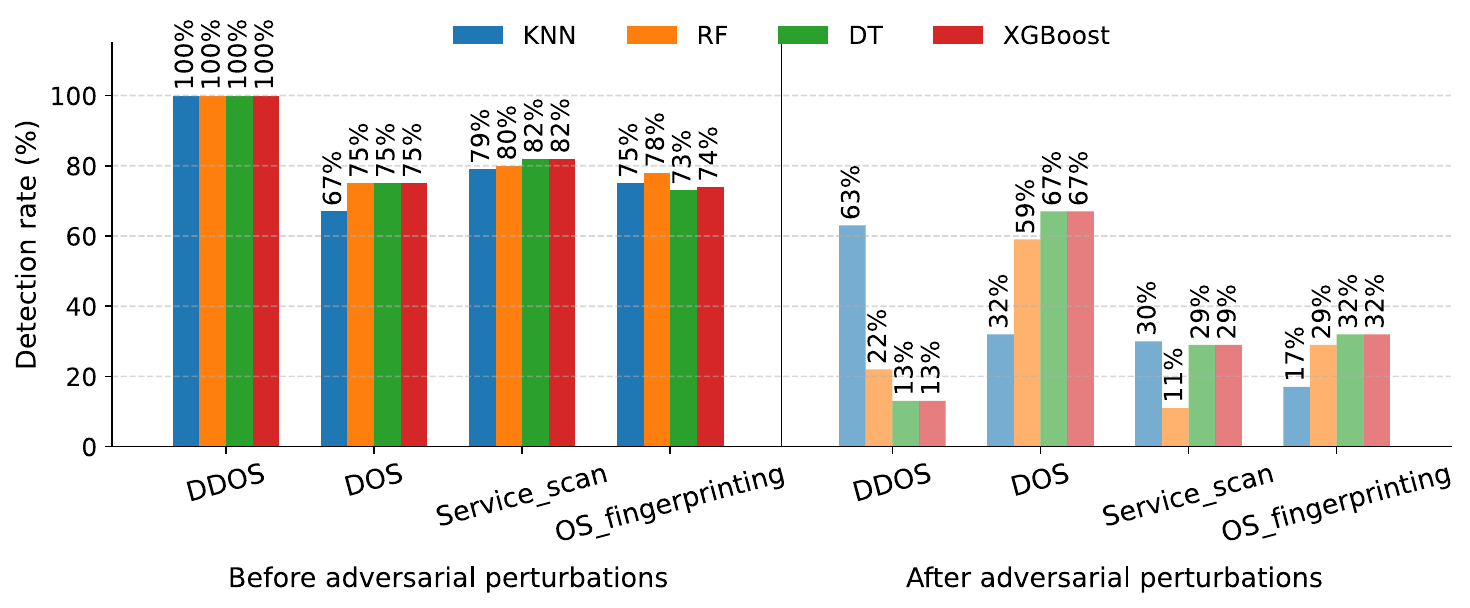}
  \caption{%
    \textbf{Impact of adversarial perturbations on IDS detection rates for the Bot-IoT dataset.}
    The bars compare detection rates of KNN, RF, DT, and XGBoost across four attack types 
    (DDOS, DOS, Service\_scan, and OS\_fingerprinting) \emph{before} and \emph{after} adversarial 
    perturbations. A marked drop in detection rate indicates the effectiveness of the adversarial 
    modifications in evading each IDS model.
  }
  \label{fig:bot_before_after}
\end{figure}

\begin{figure}[h]
  \centering
\includegraphics[width=\columnwidth]{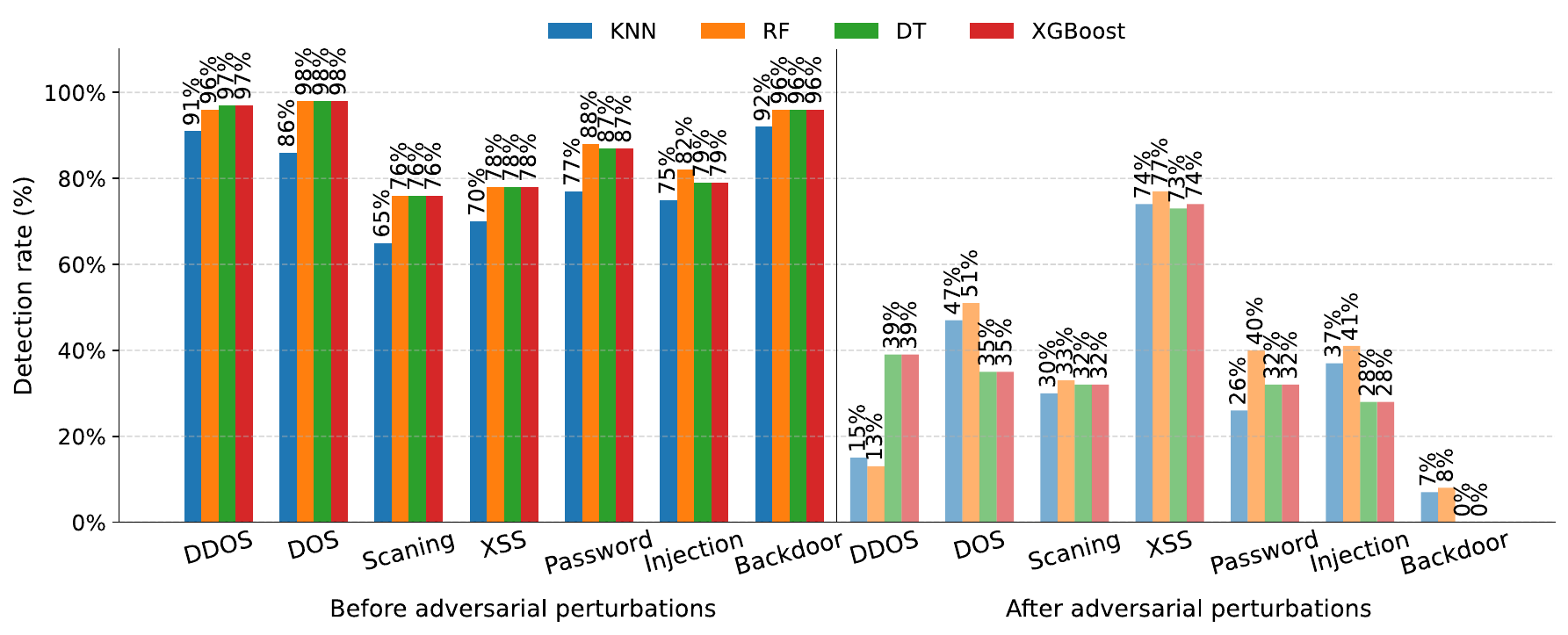}
  \caption{%
    \textbf{Impact of adversarial perturbations on IDS detection rates for the Ton-IoT dataset.}
    Similar to Figure~\ref{fig:bot_before_after}, this figure highlights how the detection 
    performance of KNN, RF, DT, and XGBoost significantly declines once adversarial samples 
    are introduced, demonstrating the robustness of our attack across multiple attack types 
    (DDOS, DOS, Scanning, XSS, Password, Injection, and Backdoor).
  }
  \label{fig:ton_before_after}
\end{figure}

Figure~\ref{fig:bot_before_after} and Figure~\ref{fig:ton_before_after} illustrate the detection performance of various models on the Bot-IoT and Ton-IoT datasets, respectively, before and after adversarial perturbations. Initially, all IDS models exhibit robust detection capabilities; however, following the introduction of adversarial samples, their effectiveness significantly declines. For example, the XGBoost detection rate for DDOS attacks drops from 100\% to 14\% on the Bot-IoT dataset and from 96\% to 38\% on the Ton-IoT dataset. Similarly, backdoor attack detection rates plummet from over 90\% to under 10\% across all models, highlighting the potency of the adversarial modifications in evading detection.

It is important to note that these adversarial instances are generated using the attacker's models and then evaluated against the defender's models. In the literature, the transferability property implies that adversarial examples crafted to evade one ML model \textit{may} also evade others; however, this outcome is not guaranteed, and some attack types might exhibit only partial transferability across different models \cite{debicha2021detect}.

This phenomenon can be attributed to several factors: some models may inherently be more robust to the specific feature manipulations used in the attack, resulting in minimal shifts in their decision boundaries; the feasible manipulation space for some attacks is limited by protocol constraints, causing the adversarial examples to remain similar to the original malicious traffic; and variations in attack complexity may also lead to negligible performance differences.

\begin{figure}[h]
  \centering
  \includegraphics[width=\columnwidth]{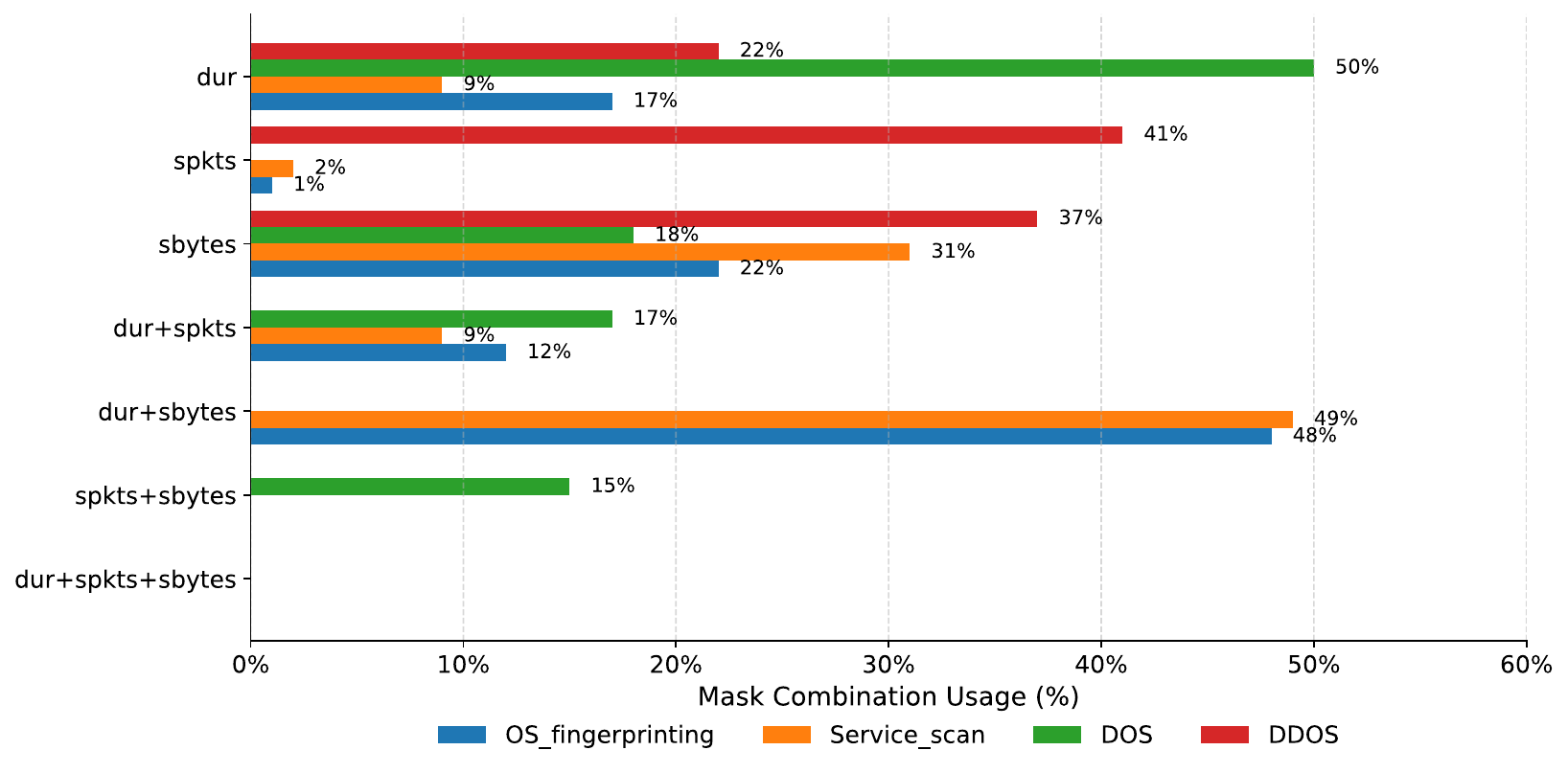}
  \caption{%
    \textbf{Percentage usage of different mask combinations for each attack in the Bot-IoT dataset.}
    Each bar represents how frequently a particular set of features (e.g., \emph{dur}, \emph{spkts}, 
    or \emph{sbytes}) is manipulated to create adversarial instances for a specific attack. 
    Higher percentages indicate that those features play a more critical role in evading detection 
    for that attack.
  }
  \label{fig:bot_mask}
\end{figure}

\begin{figure}[h]
  \centering
  \includegraphics[width=\columnwidth]{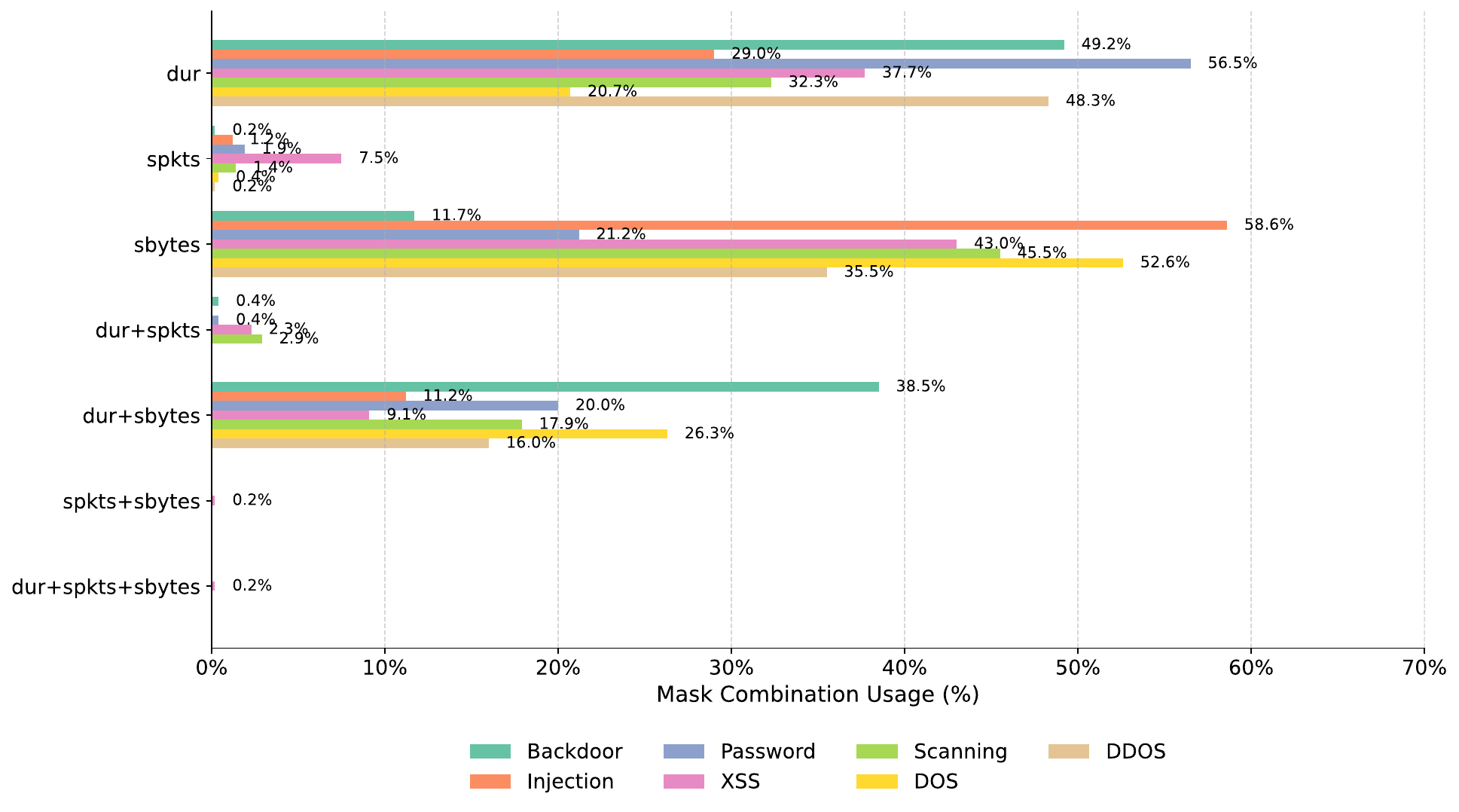}
  \caption{%
    \textbf{Proportion of mask use for every attack type in the Ton-IoT dataset.}
    Similar to Figure~\ref{fig:bot_mask}, this figure reveals which feature manipulations 
    (\emph{dur}, \emph{spkts}, \emph{sbytes}, or their combinations) are most frequently 
    employed to generate successful adversarial samples for each attack category.
  }
  \label{fig:ton_mask}
\end{figure}

\label{sec:MaskDetails}

Figures~\ref{fig:bot_mask} and~\ref{fig:ton_mask} provide a detailed breakdown of mask usage across different attack types in the Bot-IoT and Ton-IoT datasets. Our analysis reveals that masks focusing solely on the \textit{Dur} feature are predominantly employed in DoS attacks, where even minimal perturbations (often under one second) can significantly mislead the IDS. In contrast, more complex attacks - such as  backdoor or OS fingerprinting - tend to involve combinations of features (e.g., \textit{Dur+Spkts} or \textit{Dur+Sbytes}), although these multi-feature masks appear less frequently. This suggests that \textit{Dur} is particularly critical in achieving successful evasion, since minor manipulations in flow duration can lead to substantial classification errors. Conversely, certain mask combinations (e.g., \textit{spkts+sbytes}) occur infrequently, indicating that those features may have a more limited role in disrupting IDS performance for certain attacks.

\begin{figure}[h]
  \centering
 \includegraphics[width=\columnwidth]{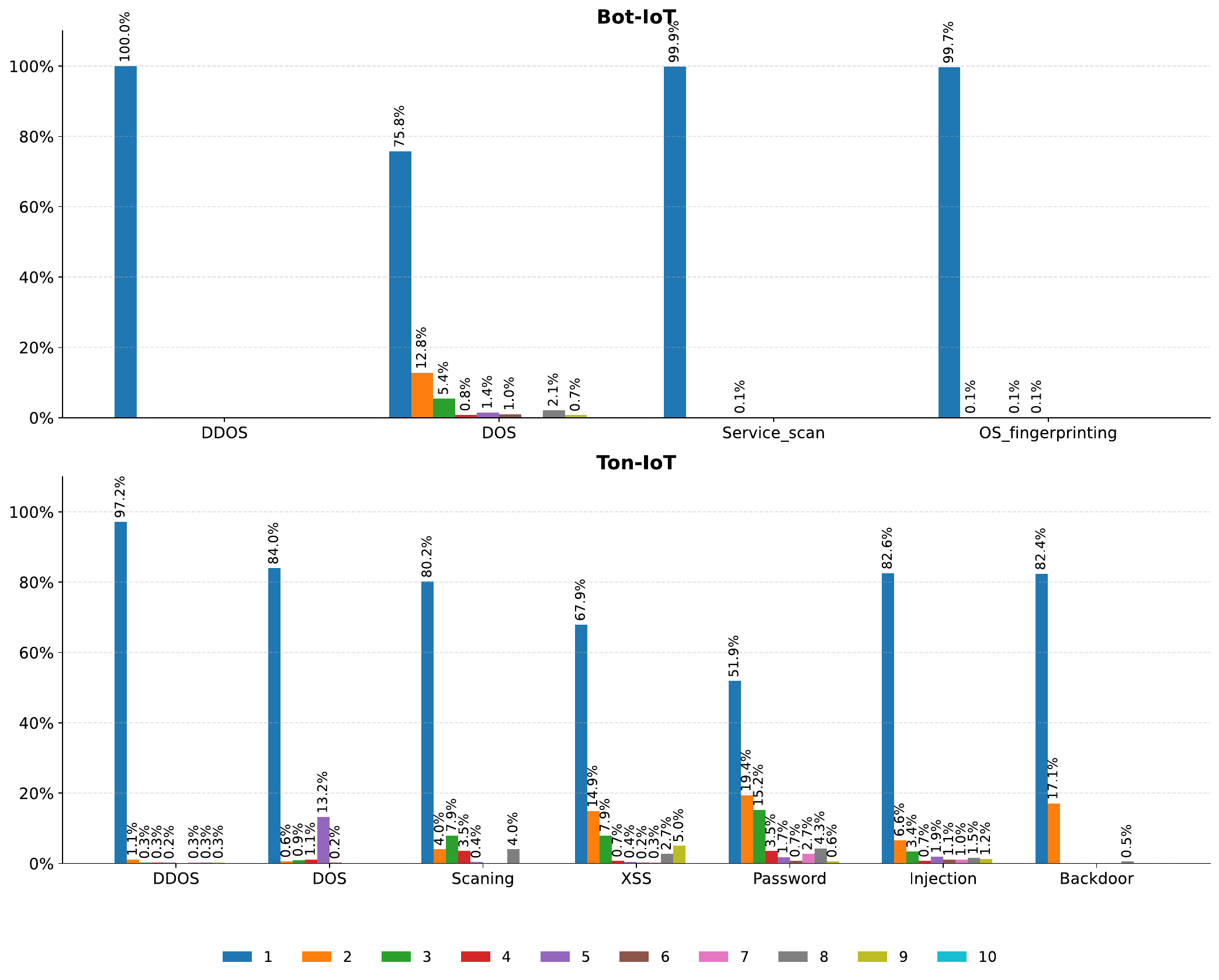}
  \caption{%
    \textbf{Number of steps required to generate successful adversarial instances across different attacks.}
    The bars represent the percentage of instances requiring each step count (1 to 10) to 
    achieve evasion. A large proportion of attacks become adversarial within the first step, 
    indicating that only minimal perturbations are necessary to fool the IDS.
  }
  \label{fig:step}
\end{figure}

\par{
Figure~\ref{fig:step} shows the percentage of adversarial instances requiring a given number of iterative perturbation steps (from 1 to 10) to evade detection for each attack. Each “step” corresponds to a single round of adding a minimal perturbation to the malicious instance and then checking whether the substitute  IDS misclassifies it. We selected an upper bound of 10 steps based on preliminary experiments indicating that most adversarial instances converge to misclassification within this range or fail to converge altogether. Notably, a significant portion of attacks achieve adversarial status within the first step, suggesting that only minimal modifications can be sufficient to mislead the IDS in certain scenarios. The figure thus highlights both the efficiency of our D2TC method in generating adversarial examples and the varying degrees of susceptibility across different attack types. 
}

\textcolor{black}{ \paragraph{Illustration of adversarial perturbations}
\label{Rev1:Q7} 
Network traffic data consists of multiple features, making it impractical to visually present every attribute's modification. To illustrate the impact of adversarial perturbations, we focus on the duration feature, a key component in network traffic analysis. Our experiments reveal that the perturbations introduced by the D2TC method remain subtle, ensuring that the modified traffic maintains its syntactic and semantic validity. For example, in the Ton-IoT dataset, the average difference in duration across various attack types ranges from 0.10 to 1.05 seconds, with DDOS attacks showing perturbations between 0.32 and 1.05 seconds. Similarly, in the Bot-IoT dataset, most models exhibit negligible differences, with a maximum perturbation of 1.17 seconds. These findings highlight that even minimal, carefully crafted changes - on the order of fractions of a second - can effectively evade IDS detection while preserving the overall structure of network traffic. This demonstrates the efficiency of our attack strategy and reinforces the need for more robust detection mechanisms capable of countering such evasive adversarial modifications.}

\begin{table}[h!]
    \centering
    \caption{Average generation time (second) of adversarial instances - Bot-IoT}
    \label{tab:bot-iot}

    \begin{tabularx}{0.8\textwidth}{LCCCC}
        \toprule
        \textbf{Attack Type} & \textbf{KNN} & \textbf{RF} & \textbf{DT} & \textbf{XGBoost} \\
        \midrule
        DDOS             & 0.05 & 0.11 & 0.01 & 0.02 \\
        DOS              & 0.02 & 1.40 & 1.29 & 1.53 \\
        Service Scan     & 0.07 & 0.22 & 0.01 & 0.02 \\
        OS Fingerprinting & 0.07 & 0.27 & 0.01 & 0.02 \\
        \bottomrule
    \end{tabularx}
\end{table}

\begin{table}[h!]
    \centering
    \caption{Average generation time (second) of adversarial instances - Ton-IoT}
    \label{tab:ton-iot}

    \begin{tabularx}{0.8\textwidth}{LCCCC}
        \toprule
        \textbf{Attack Type} & \textbf{KNN} & \textbf{RF} & \textbf{DT} & \textbf{XGBoost} \\
        \midrule
        DDOS       & 1.47 & 0.40 & 0.01 & 0.09 \\
        DOS        & 0.08 & 0.21 & 0.01 & 0.74 \\
        Scanning   & 0.06 & 0.24 & 0.01 & 0.12 \\
        XSS        & 1.02 & 1.62 & 0.03 & 0.85 \\
        Password   & 0.43 & 1.85 & 0.02 & 0.40 \\
        Injection  & 1.65 & 1.88 & 0.08 & 1.25 \\
        Backdoor   & 0.08 & 0.29 & 0.00 & 0.08 \\
        \bottomrule
    \end{tabularx}
\end{table}

\paragraph{Computational efficiency of our adversarial instance generation process}
\label{Rev1:Q5} 
In addition to the impact on detection performance, we also report on the computational efficiency of our adversarial instance generation process. Tables~\ref{tab:bot-iot} and~\ref{tab:ton-iot} summarize the average generation times for adversarial instances across various attack types and ML models on the Bot-IoT and Ton-IoT datasets, respectively. For example, on the Bot-IoT dataset, the average generation time for DDOS attacks is as low as 0.05 seconds for KNN and 0.11 seconds for RF. On the Ton-IoT dataset, generation times range from 0.01 seconds (DT) to 1.65 seconds (KNN), depending on the attack type and ML model. It should be noticed that the training process is executed offline as a one-time operation, ensuring that it does not affect real-time performance. Moreover, the testing phase exhibits linear time complexity relative to the number of samples, making the testing time negligible compared to the generation time. These findings confirm that our approach is computationally efficient and well-suited for deployment in real-time IDS environments.

\paragraph{Comparison with other adversarial attack methods}
\label{para:CompWithOAdvAtt}

In addition to the proposed D2TC method, recent literature has explored alternative adversarial attack approaches, notably on-manifold attacks and generative attacks. On-manifold transferability attacks exploit the intrinsic structure of the data by operating directly on the manifold of legitimate traffic samples. These methods generate adversarial examples that lie close to the data distribution, ensuring high semantic similarity with genuine instances (e.g., Xiao, Jiancong, et al. \cite{xiao2025understanding}). However, such approaches typically require a comprehensive understanding of the underlying data manifold, which may not be feasible in black-box scenarios where the attacker’s knowledge is limited. In contrast, generative attacks - often implemented through generative adversarial networks (GANs) - aim to learn the distribution of normal network traffic to generate adversarial samples that are virtually indistinguishable from real data (e.g.,  Dong et al. \cite{dong2025masqueradegan} and Zhao et al. \cite{zhao2021attackgan}). While these generative methods can produce highly realistic adversarial examples, they generally entail significant computational complexity and demand large amounts of training data to capture the nuances of network traffic. Our D2TC approach addresses these challenges by directly manipulating key network features within predefined semantic and syntactic constraints, thereby ensuring both the validity of the generated adversarial samples and their effective transferability to black-box IDS models. This makes the D2TC method a practical alternative, especially in environments where the attacker has limited knowledge of the target system and computational resources are constrained.

\subsection{Defense Effectiveness}
\label{sec:DefEffec}

\begin{figure}[h]
\centering
\includegraphics[width=\columnwidth]{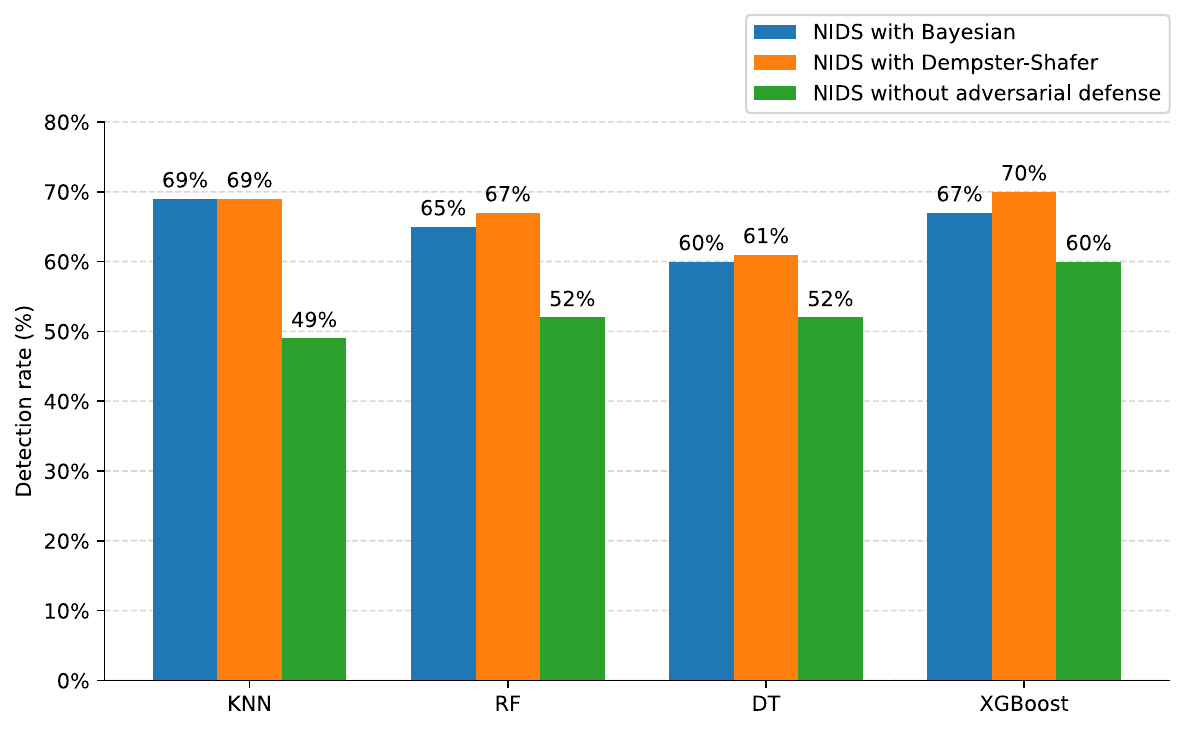}
\caption{Defender’s NIDS performance under adversarial traffic, with and without our proposed defense (using Bayesian and Dempster-Shafer fusion).}
\label{defnse with and without diagram}
\end{figure}

To evaluate the effectiveness of our proposed adversarial defense, we integrated it into our threat scenario and launched two attack rounds. First, the defender’s NIDS operated without any protection. Second, we equipped the defender’s NIDS with our adversarial defense, using both Bayesian and Dempster-Shafer (DS) fusion rules. Tables~\ref{tab:nids_no_defense}--\ref{tab:nids_dempster} present the detection rates of the defender’s NIDS in each setup, while Figure~\ref{defnse with and without diagram} provide a visual summary.

\begin{table}[h]
    \centering
    \caption{Detection rates of the defender's NIDS without adversarial defense across four attacker substitute models (KNN, RF, DT, XGBoost). The final column shows the average detection rate.}
    \label{tab:nids_no_defense}
    \begin{tabularx}{0.8\textwidth}{l *{5}{>{\centering\arraybackslash}X}}
        \toprule
        \multirow{2}{*}{\textbf{Defender models}} & \multicolumn{4}{c}{\textbf{Attacker's substitute models}} & \multirow{2}{*}{\textbf{Average}} \\
        \cmidrule(lr){2-5}
         & \textbf{KNN} & \textbf{RF} & \textbf{DT} & \textbf{XGBoost} &  \\
        \midrule
        KNN      & 49\% & 47\% & 50\% & 51\% & 49\% \\
        RF       & 56\% & 49\% & 53\% & 52\% & 53\% \\
        DT       & 55\% & 50\% & 52\% & 52\% & 52\% \\
        XGBoost  & 69\% & 53\% & 69\% & 63\% & 64\% \\
        \bottomrule
    \end{tabularx}
\end{table}

\begin{table}[h]
    \centering
    \caption{Detection rates of the defender's NIDS with our adversarial defense (Bayesian fusion) against four attacker substitute models. The final column shows the average detection rate.}
    \label{tab:nids_bayesian}
    \begin{tabularx}{0.8\textwidth}{l *{5}{>{\centering\arraybackslash}X}}
        \toprule
        \multirow{2}{*}{\textbf{Defender models}} & \multicolumn{4}{c}{\textbf{Attacker's substitute models}} & \multirow{2}{*}{\textbf{Average}} \\
        \cmidrule(lr){2-5}
         & \textbf{KNN} & \textbf{RF} & \textbf{DT} & \textbf{XGBoost} &  \\
        \midrule
        KNN      & 65\% & 61\% & 74\% & 71\% & 68\% \\
        RF       & 76\% & 63\% & 69\% & 58\% & 67\% \\
        DT       & 66\% & 39\% & 68\% & 68\% & 60\% \\
        XGBoost  & 79\% & 56\% & 74\% & 72\% & 70\% \\
        \bottomrule
    \end{tabularx}
\end{table}

\begin{table}[h]
    \centering
    \caption{Detection rates of the defender's NIDS with our adversarial defense (Dempster-Shafer fusion) against four attacker substitute models. The final column shows the average detection rate.}
    \label{tab:nids_dempster}
    \begin{tabularx}{0.8\textwidth}{l *{5}{>{\centering\arraybackslash}X}}
        \toprule
        \multirow{2}{*}{\textbf{Defender models}} & \multicolumn{4}{c}{\textbf{Attacker's substitute models}} & \multirow{2}{*}{\textbf{Average}} \\
        \cmidrule(lr){2-5}
         & \textbf{KNN} & \textbf{RF} & \textbf{DT} & \textbf{XGBoost} &  \\
        \midrule
        KNN      & 65\% & 61\% & 75\% & 71\% & 68\% \\
        RF       & 76\% & 63\% & 69\% & 71\% & 70\% \\
        DT       & 66\% & 56\% & 76\% & 60\% & 63\% \\
        XGBoost  & 82\% & 66\% & 76\% & 73\% & 74\% \\
        \bottomrule
    \end{tabularx}
\end{table}

\paragraph{Overall Results}
Table~\ref{tab:nids_no_defense} illustrates that, without any adversarial defense, the NIDS performance can drop to around 50\% on average, underscoring the potency of the adversarial samples in evading detection. In contrast, Tables~\ref{tab:nids_bayesian} and~\ref{tab:nids_dempster} show a notable improvement - often exceeding 20\% - when our defense is employed. This boost highlights the value of our ensemble-based approach, which filters out adversarial traffic before it reaches the NIDS.

\paragraph{Comparison of Bayesian vs. Dempster-Shafer Fusion}
\label{sec:CompBayDSFusion}
As seen in Figure~\ref{defnse with and without diagram} (and Tables~\ref{tab:nids_bayesian} and~\ref{tab:nids_dempster}), there is sometimes only a small difference between the performance of the Bayesian-fused NIDS and the Dempster-Shafer-fused NIDS. Both methods rely on the same sub-detector reliability weighting scheme and differ primarily in how they aggregate final decisions. Consequently, they often yield comparable detection rates across KNN, RF, and DT. For XGBoost, however, we observe a 10\% difference in performance when no adversarial defense is used, indicating that XGBoost can be more robust under certain conditions but remains susceptible to well-crafted perturbations.

\subsection{Comparison with State-of-the-Art Adversarial Defenses}
\label{subsec:CompWithSOTAAdvDef}

Various defenses against evasion attacks have been explored in the literature, but many suffer from critical drawbacks~\cite{he2023adversarial}. Adversarial training, a widely used defense strategy \cite{xiong2023aidtf, rashid2022adversarial}, improves model robustness by incorporating adversarial examples during training. However, it often leads to reduced accuracy on benign inputs and demands substantial computational resources~\cite{he2023adversarial, debicha2021adversarial}. Similarly, Input pre-processing techniques \cite{qiu2021efficient, meng2017magnet} can compromise detection performance on normal traffic\cite{he2023adversarial, debicha2023review}. In contrast, our proposed defense preserves the original NIDS parameters and performance on clean inputs while significantly enhancing resilience to adversarial attacks. By incorporating reliability weights into an ensemble framework - leveraging Bayesian and Dempster-Shafer fusion rules - our approach dynamically adapts to different adversarial strategies, maintaining effectiveness even in black-box attack scenarios while ensuring operational efficiency.

We compare our proposed defense against a state-of-the-art adversarial detection method used frequently in the literature~\cite{ grosse2017statistical, kumar2025nids, wang2022manda, debicha2021detect}. Adversarial detection involves training a classifier specifically on adversarial instances, allowing it to distinguish adversarial from normal traffic. This method aligns with our defense strategy but lacks the adaptive fusion mechanisms employed in our approach.

\begin{figure}[h]
\centering
\includegraphics[width=\columnwidth]{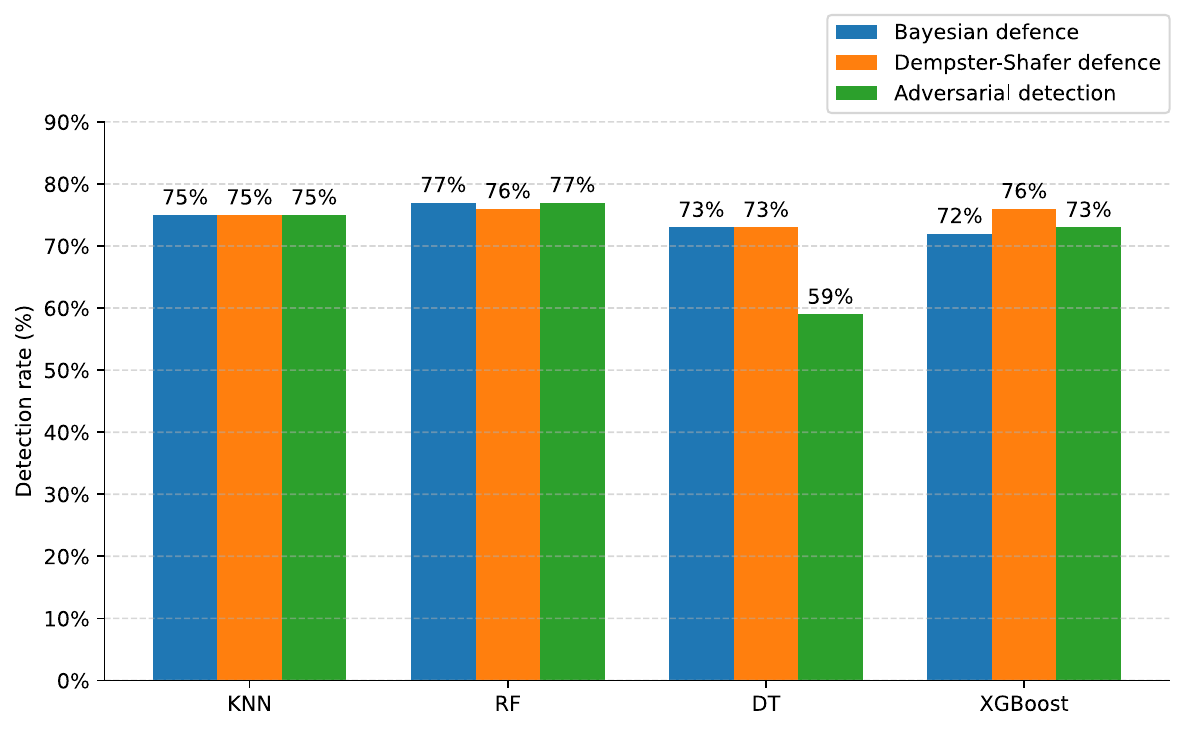}
\caption{Comparison of our Bayesian and Dempster-Shafer fusion-based defenses with a state-of-the-art adversarial detection method across various ML models}
\label{comparison and state of the art}
\end{figure}

Figure~\ref{comparison and state of the art} illustrates the comparative effectiveness of our defense and adversarial detection across various machine learning models. While the performance gap is small for most models (KNN, RF, and XGBoost), our Bayesian and Dempster-Shafer fusion-based defenses exhibit a notable advantage for DT, where they significantly outperform the baseline. However, it is important to note that even with these improvements, the overall detection rate in most configurations remains below 80\%, underscoring the persistent challenge of defending against sophisticated adversarial attacks in black-box environments.

\paragraph{Interpretation and Insights.}
While our defense does not drastically alter the underlying NIDS parameters or training procedure - thus maintaining the original detection performance on benign traffic - it substantially improves resilience to adversarial samples. The slight differences between Bayesian and DS fusion reflect how final decisions are aggregated, rather than major changes in sub-detector reliability weighting. Moreover, variations among different ML models (KNN, RF, DT, XGBoost) highlight that certain algorithms exhibit more sensitivity to adversarial perturbations than others, emphasizing the importance of ensemble approaches that leverage multiple perspectives on adversarial traffic.

Overall, these findings confirm that our proposed defense mechanism effectively mitigates adversarial attacks without sacrificing operational efficiency, thereby offering a robust and practical solution for network intrusion detection in IoT environments.

\subsection{Discussion and Future Directions}
\label{sec:Discussion}

The experimental results provide several key insights into both the vulnerabilities of ML-based IDSs in IoT networks and the effectiveness of our proposed adversarial attack and defense mechanisms. Our evaluation demonstrates that even minimal adversarial perturbations – achieved through the D2TC method – can significantly degrade IDS performance, with detection rates for some attacks dropping by as much as 80\%. In addition, the high transferability of adversarial examples generated on substitute models underscores the feasibility of black-box attacks, where an attacker with limited knowledge can craft perturbations that remain effective against the defender's IDS.

These adversarial examples not only reveal critical vulnerabilities in current systems but also have practical implications. Even subtle modifications – on the order of fractions of a second in the Duration feature – can cause drastic misclassifications, enabling attackers to bypass network security with minimal effort. Moreover, these adversarial samples serve as invaluable tools for red-team exercises, offering realistic benchmarks to evaluate IDS performance under sophisticated attack scenarios. Insights gained from this analysis can inform the development of advanced defense strategies, such as incorporating adversarial training or integrating ensemble-based detection methods.

On the defense side, our proposed adversarial detection mechanism – integrating ensemble methods with Bayesian and Dempster-Shafer fusion rules – demonstrated a significant improvement in resilience, recovering detection rates by more than 20\% compared to the baseline. While both fusion methods yield comparable overall performance, certain models (e.g., decision trees) benefit more from the fusion-based approach, highlighting the importance of properly calibrated reliability weights.

Despite these promising outcomes, several limitations remain. First, the D2TC method relies on manipulating specific network features (e.g., \emph{Dur}, \emph{Spkts}, and \emph{Sbytes}), which may limit its applicability to other network traffic types or IDS architectures that utilize different feature sets. Second, our defense mechanism, which involves training an ensemble of sub-detectors, incurs additional computational overhead compared to simpler approaches such as adversarial training or feature filtering. 

\color{black}
To assess the impact of this overhead, we conducted a comparative analysis of the execution time during the detection phase (inference) between our fusion-based scheme and standard flat classifiers (e.g., a single XGBoost or Random Forest model). Our measurements show that the complete defense pipeline, including feature extraction, parallel sub-detector evaluation, reliability weighting, and final decision fusion, introduces an average inference delay approximately 2.3 times greater than using a single classifier. It is important to emphasize that this additional delay remains within acceptable limits for real-time intrusion detection systems. Furthermore, because the sub-detectors operate independently, the architecture is inherently parallelizable, allowing further performance improvements through hardware acceleration or lightweight GPU offloading. In summary, although the ensemble-based architecture introduces moderate computational overhead, it does not significantly affect the real-time responsiveness of the defense.

\color{black}
Moreover, while our experiments on the Bot-IoT and Ton-IoT datasets demonstrate the effectiveness of our approach in controlled environments, further research is needed to assess its real-time performance and scalability to larger, more diverse datasets.

Future work should explore the adaptability of the D2TC method across various network conditions and investigate hybrid defense strategies that combine our ensemble-based approach with robust feature selection and dynamic adversarial training. Such efforts would contribute to the development of a more comprehensive and resilient defense framework, better capable of mitigating adversarial threats in evolving network environments.

Overall, our findings reveal that current ML-based IDSs are highly susceptible to carefully crafted adversarial attacks, underscoring the necessity for robust, efficient defense strategies – such as the one proposed in this work – to enhance the security of IoT networks.

\color{black}

\section{Conclusion}
\label{sec:conclusion}

The potential of using NIDS based on machine learning algorithms raises intriguing security issues. Indeed, despite their impressive performance, these ML models are prone to various types of adversarial attacks, especially evasion attacks. Due to their prevalence and feasibility among all adversarial attacks, evasion attacks were considered in this paper to generate adversarial traffic capable of evading the intrusion detection system. This work was performed in the context of IoT networks, which represent highly constrained environments.

The proposed framework includes two main contributions. The first is a realistic adversarial algorithm capable of generating valid adversarial network traffic by adding small perturbations, thus evading NIDS protection with high probability while maintaining the underlying logic of the network attack.

The second component of the proposed framework is a reactive defense that limits the impact of the proposed attack. This defense, inspired by adversarial detection, capitalizes on the fact that it does not change the initial performance of the NIDS since it provides an additional layer of security independent of the model. The proposed defense is considered modular because it uses an ensemble method called bagging. In addition to this ensemble method, it also includes a contextual discounting method that improves the overall performance of the defense.

The results demonstrated the efficacy of our adversarial algorithm in generating traffic that successfully evaded the intrusion detection system, underscoring the vulnerability of machine learning-based NIDS to sophisticated evasion techniques. Conversely, our proposed defense proved effective in detecting the majority of adversarial traffic, exhibiting promising performance compared to state-of-the-art defenses.  Since the proposed framework is easily adaptable to other domains, evaluating its performance in other highly constrained domains would be an interesting future work.

\section*{Acknowledgement}
The authors would like to express their sincere gratitude to \textit{Mohammed Cherif Tifoura} for his meticulous proofreading of the manuscript. His valuable insights and suggestions have significantly improved the clarity and quality of this work.

\section*{Declaration of competing interest}
The authors declare that they have no known competing financial interests or personal relationships that could have appeared to influence the work reported in this paper.
\bibliography{sn-bibliography}

\end{document}